\documentclass[
  aps,
  prl,
  reprint,
  floatfix,
  superscriptaddress,
  longbibliography,
  amsmath,
  amssymb
]{revtex4-2}

\usepackage[version=3]{mhchem}
\usepackage[T1]{fontenc}
\usepackage{bm}
\usepackage{graphicx}
\usepackage{xcolor}
\usepackage{booktabs}
\usepackage{microtype}
\usepackage{placeins}
\usepackage[hidelinks]{hyperref}

\hypersetup{
  pdftitle={Geometry-Dependent Nonlocal Valence Screening Following Core Ionization of the Water Dimer},
  pdfauthor={Vibin Abraham and Bo Peng},
  pdfsubject={Correlated core-hole screening in a hydrogen-bonded water dimer},
  pdfkeywords={core ionization, water dimer, nonlocal screening, real-time coupled-cluster theory, X-ray photoelectron spectroscopy}
}
\begin{document}

\preprint{arXiv preprint}
\title{Geometry-Dependent Nonlocal Valence Screening Following Core Ionization of the Water Dimer}

\author{Vibin Abraham}
\email{vibin.abraham@pnnl.gov}
\affiliation{Integrated Discovery Sciences Directorate, Pacific Northwest National Laboratory, Richland, Washington 99354, USA}
\author{Bo Peng}
\thanks{Corresponding author: \href{mailto:peng398@pnnl.gov}{peng398@pnnl.gov}}
\affiliation{Integrated Discovery Sciences Directorate, Pacific Northwest National Laboratory, Richland, Washington 99354, USA}
\date{August 2026}

\begin{abstract}
Core ionization is spatially localized, but the correlated valence response
that screens the resulting hole need not be. Using the water dimer as a
minimal hydrogen-bonded system, we ask whether acceptor-site O~1s ionization
recruits valence channels on the neighboring donor molecule and how proton
displacement redistributes that response. We introduce correlated
shifted-start real-time $\Lambda$-coupled-cluster theory to connect the
core-hole spectrum with orbital- and fragment-resolved screening pathways. At
equilibrium, the satellite response contains donor-local and intermolecular
charge-transfer-like contributions. Their intermolecular origin is supported
by the strong suppression of both contributions when the monomers are
separated. In a fixed-nuclei geometry scan, elongating the hydrogen-bonded
donor O--H bond nearly doubles their combined share of the satellite response,
while analysis of the first approximately 10~fs of the fixed-nuclei electronic
response reveals stronger modulation involving a donor-localized valence
orbital. More broadly, this work shows how localized core ionization can serve
as a site-selective probe of electronic communication, with satellite
structure revealing how nuclear geometry redirects many-electron screening
through molecular environments.
\end{abstract}

\maketitle

\begin{figure*}
  \centering
  \includegraphics[width=0.98\textwidth]{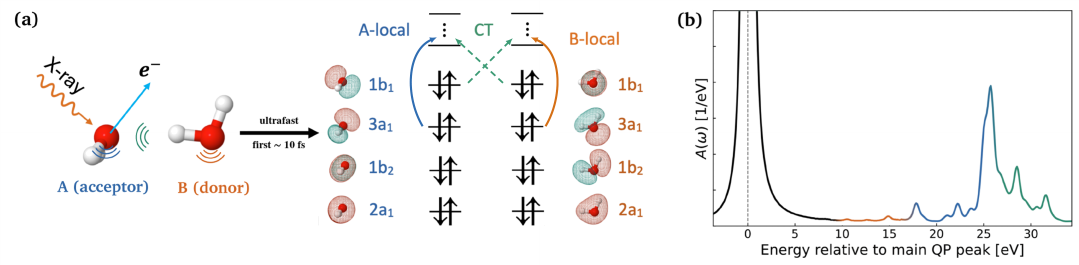}
  \caption{Physical question and spectral signature of nonlocal core-hole
  screening in the water dimer. (a) An incident X-ray ejects an O~1s
  photoelectron from molecule $A$, the hydrogen-bond acceptor, and launches
  correlated valence relaxation during the first approximately 10~fs, the
  fixed-nuclei electronic regime examined here. The
  neighboring molecule $B$ is the hydrogen-bond donor. The concentric
  three-arc symbols represent electronic, not nuclear, motion. Blue and orange
  denote local occupied-to-virtual electron transitions on $A$ ($A$-local) and $B$ ($B$-local),
  respectively, whereas green denotes an intermolecular charge-transfer-like (CT) electron
  transition. (b) Equilibrium correlated shifted-start real-time $\Lambda$-coupled-cluster singles-and-doubles
  (RT-$\Lambda$CCSD) O~1s spectrum relative to the main quasiparticle peak, whose position is
  $-543.562$~eV. The full 10--30~eV satellite manifold carries the relevant
  information: the weak features below approximately 16~eV are predominantly
  donor-local, acceptor-local shake-up dominates the 18--25~eV region, and
  CT-like character increases above approximately
  26~eV. The low-energy features are especially diagnostic because their
  leading configurations reside on the initially non-ionized donor molecule
  and therefore provide the most direct spectral signature that
  screening has crossed the hydrogen bond. The curve color is a coarse guide
  to these regimes: black for the quasiparticle and leading region, orange for
  the donor-local-dominant region, blue for the acceptor-local-dominant region,
  and green where CT-like character increases. The blended boundaries are
  approximate and do not represent a pointwise channel decomposition.}
  \label{fig:equilibrium_overview}
\end{figure*}

\section{Introduction}

{
Removing a deep core electron abruptly changes the potential experienced by the remaining electrons. The resulting correlated ionized wavepacket relaxes before appreciable nuclear motion through orbital screening and shake-up processes that generate the quasiparticle and satellite structure of an X-ray photoelectron spectrum~\cite{greczynski2020x,Golze2020,van2018assessing,cederbaum1977complete,conradie2022xps,klevak2014charge,bagus2022origin,ghiasi2019charge}. Because the perturbation begins at a specific atomic site, core ionization offers a direct way to ask how far the ensuing many-electron response extends. In an interacting molecular environment, the central question is therefore not only how the ionized molecule screens its core hole, but whether that localized perturbation recruits valence channels on a neighboring molecule.
}

{
Hydrogen bonding makes this question especially consequential because it couples molecules while preserving chemically identifiable donor and acceptor fragments. It reorganizes occupied and virtual orbitals and can mediate intermolecular charge redistribution, relaxation, and energy transfer~\cite{garrett2005role,mherbert2017hydrated,guo2002xes_water,tokushima2008two_motifs,fuchs2008isotope_temperature,slavicek2014proton,fransson2016xray,takahashi2022xes_interpretation}. In water, attosecond measurements have resolved orbital-dependent photoionization dynamics in the isolated molecule~\cite{huppert2016attosecond}. All-X-ray transient absorption further revealed a subfemtosecond response in liquid water following valence ionization, during which hydrogen motion remains effectively frozen~\cite{li2024attosecond_water}. This experimentally accessible separation of electronic and nuclear timescales motivates asking how a hydrogen-bonded environment distributes the prompt electronic response. Cluster studies have also established intermolecular Coulombic decay, coherent charge motion, and proton-transfer-coupled ionization dynamics~\cite{stoychev2010water_dimer_icd,richter2018competition_icd_pt,Periyasamy2009WaterDimer,kamarchik2010water_dimer,hartweg2021water_dimer,kumar2024relaxation_2a1_dimer,yamazoe2019measurements,inhester2019spectroscopic,vila2022water_dimer_xps,felicissimo2005water_dimer_core}. These observations show that ionization-driven response in hydrogen-bonded water can be nonlocal. They do not, however, identify how a localized core hole distributes correlated screening among acceptor-local, donor-local, and intermolecular valence channels, or how that distribution changes with the proton coordinate.
}

{
The water dimer is the minimal controlled system in which to separate these pathways. Its single donor--acceptor hydrogen bond, two distinguishable fragments, and site-resolved O~1s orbitals allow us to create a core hole on the acceptor and track the initially non-ionized donor independently. We can then classify the satellite component area into acceptor-local, donor-local, and charge-transfer-like channels, while a fixed-geometry separation scan tests whether the latter two require intermolecular coupling. Because the dimer is not a surrogate for bulk water, we use it instead as a molecular testbed to isolate the spatial origin and geometry dependence of nonlocal core-hole screening.
}

{
This reduction also provides a controlled reference for more complex hydrogen-bonded systems. Combining site-selective core ionization, fragment-resolved satellite analysis, and fixed-geometry scans distinguishes local relaxation from environment-mediated screening and suggests which coordinates may be especially sensitive to that redistribution. The dimer can therefore benchmark real-time and frequency-domain many-body theories before researchers extend the same questions to larger water clusters, solvated molecules, or hydrogen-bonded materials.
}

{
Existing approaches capture complementary parts of this problem but leave a gap between scalable dynamics and many-body spectral interpretation. Researchers have applied real-time time-dependent density functional theory (TDDFT) to ionized water aggregates, proton-transfer and inner-valence relaxation, electronic stopping, and intermolecular Coulombic decay~\cite{chalabala2018rttddft_water_dimer,reeves2017proton_water,yao2019kshell_water,sharma2020rttddft_water_chains,wang2024nonhermitian_icd,wang2025rttddft_icd_water_dimer}. Its Kohn--Sham density-response framework provides direct access to charge and nuclear dynamics at useful system sizes. However, it does not naturally provide a systematic quasiparticle--satellite balance or assign satellite intensity to correlated shake-up configurations, and both outputs remain sensitive to the exchange--correlation approximation. Conversely, frequency-domain approaches based on algebraic-diagrammatic construction, symmetry-adapted-cluster configuration interaction (SAC-CI), $GW$/cumulant, and coupled-cluster (CC) Green's functions or equation-of-motion theory provide correlated ionization energies and satellite structure~\cite{schirmer1983extended,dreuw2015algebraic,ahmed2025core,nakatsuji1991cluster,ohtsuka2006inner,golze2019gw_compendium,kas2014cumulant,guzzo2011valence,ammar2024cumulant,kocklaeuner2025gw,coriani2015communication,ranga2021core,faber2019xes_cc,nanda2020rixs,nooijen1992coupled,nooijen1993coupled,bhaskaran2016coupled,peng2018green,shee2019coupled,backhouse2022constructing}. Prior real-time equation-of-motion coupled-cluster (RT-EOM-CC) and CC Green's-function calculations on the water dimer further followed ionization potentials, satellite positions, and their geometry dependence~\cite{vila2022water_dimer_xps}. Those calculations established access to many-body spectral features, but they did not partition the evolving core-hole response into acceptor-local, donor-local, and intermolecular screening pathways.
}

{
For hydrogen-bonded core ionization, we therefore need a correlated framework that treats the main quasiparticle and satellite manifold together while exposing the microscopic screening components in real time. Time-dependent coupled-cluster developments bridge the existing viewpoints through bivariational, orbital-adaptive, real-time EOM, linear/transient-response, and reduced-scaling formulations~\cite{kvaal2012abinitio,nascimento2016linear,pedersen2019symplectic,koulias2019relativistic,kristiansen2020stability,skeidsvoll2020transient,pedersen2021interpretation,wang2022accelerating,peyton2023reduced,wang2025rtcc3}. In particular, RT-EOM-CC cumulant theory propagates a core-ionized $(N-1)$-electron state and generates its Green's function directly~\cite{schonhammer1978time,rehr2020equation,vila2022rteom_ccsd,vila2024rt,vila2025efficient,pathak2023real}. Its original single-exponential form starts from an uncorrelated core-hole determinant, however, and therefore omits correlation inherited from the neutral state at the instant of sudden ionization~\cite{peng2024exploring}. Double-coupled-cluster (dCC) formulations retain that coupling through neutral- and ionized-sector exponentials, with approximate time-dependent variants (TD-dCC) reducing the resulting complexity~\cite{peng2024exploring,abraham2026elucidating}. This trade-off motivates a formulation that retains single-exponential propagation while incorporating the dominant neutral-state correlation into the initial ionized wavepacket and its spectral projection.
}

{
We therefore introduce correlated shifted-start real-time $\Lambda$-coupled-cluster theory (RT-$\Lambda$CC) for the core-hole Green's function. The ionized-sector amplitudes begin from the correlated sudden-ionization state, and an explicit $\Lambda$-weighted overlap with the neutral coupled-cluster left state evaluates the projection needed for the spectral response. RT-$\Lambda$CC consequently preserves single-exponential propagation while recovering the leading neutral-sector effects seen in the more elaborate dCC description. For the present chemical question, this goes beyond a peak-only calculation. The correlated Green's function retains the quasiparticle--satellite balance, while the same overlap decomposes satellite intensity into orbital- and fragment-resolved screening contributions~\cite{peng2024exploring,abraham2026elucidating,krishnan2026chemical}.
}

{
This framework lets us test a causal sequence. We first benchmark RT-$\Lambda$CC against the exact impurity-model spectrum and show that it follows the leading behavior of the corresponding dCC propagation. We then ask whether acceptor O~1s ionization produces a donor-local or interfragment satellite response at equilibrium, and use monomer separation to test whether that response requires intermolecular coupling. Finally, we vary the donor O--H coordinate at fixed nuclei to determine how proton displacement redistributes the screening pathways. Separating the monomers suppresses the donor-involving response, whereas elongating the donor O--H bond enhances both donor-local and charge-transfer-like contributions. An orbital-resolved short-time analysis then identifies which donor-indexed excitation-amplitude group carries this redistribution and when it is activated. Figure~\ref{fig:equilibrium_overview} summarizes the site-selective perturbation, fragment-resolved channels, and equilibrium spectral signature. Together, these tests connect a localized core-hole perturbation to geometry-dependent screening and the satellite structure through which we observe it.
}

\section{Theory}

\subsection{Correlated sudden-ionization state}

The exact diagonal retarded one-particle Green's function for a core
spin-orbital $c$ separates into electron-addition and electron-removal
components, $G_c^R(t)=G_{c,+}^R(t)+G_{c,-}^R(t)$. Photoemission probes the
electron-removal component, which has the exact sudden-ionization form
\begin{equation}
G_{c,-}^{R}(t)
=
-i\Theta(t)e^{-iE_g^{(N)}t}
\left\langle
\Psi_c^{(N-1)}
\middle|
e^{iHt}
\middle|
\Psi_c^{(N-1)}
\right\rangle ,
\label{eq:sudden_green}
\end{equation}
with
\begin{equation}
\left|\Psi_c^{(N-1)}\right\rangle
=
a_c\left|\Psi_g^{(N)}\right\rangle .
\label{eq:sudden_state}
\end{equation}
For brevity, we denote $G_{c,-}^{R}(t)$ by $G_c^R(t)$ below.
Equation~\eqref{eq:sudden_green} is therefore exact within the
electron-removal sector. Because it describes only electron removal, the full
retarded Green's function additionally contains $G_{c,+}^R(t)$. The
coupled-cluster parametrization and its CCSD
truncation introduce the approximations used in the calculations.
Within coupled-cluster theory,
\begin{equation}
\begin{aligned}
\left|\Psi_g^{(N)}\right\rangle
&=e^{T^{(N)}}\left|\phi_0^{(N)}\right\rangle ,\\
\left\langle\Psi_g^{(N)}\right|
&=\left\langle\phi_0^{(N)}\right|
(1+\Lambda^{(N)})e^{-T^{(N)}} .
\end{aligned}
\label{eq:cc_ground}
\end{equation}
Because the neutral excitation operator commutes with removal from a fully occupied core orbital, the suddenly ionized state inherits the neutral-state correlation,
\begin{equation}
\left|\Psi_c^{(N-1)}\right\rangle
=
e^{T^{(N)}}\left|\phi_c^{(N-1)}\right\rangle ,
\qquad
\left|\phi_c^{(N-1)}\right\rangle
=
a_c\left|\phi_0^{(N)}\right\rangle .
\label{eq:cc_sudden_state}
\end{equation}
Section~S2.1 of the Supporting Information gives the same exact
electron-removal formulation as Eq.~\eqref{eq:sudden_green} and derives the
reduction of the sudden-ionization state and the resulting shifted initial
condition.

\subsection{Correlated shifted-start \texorpdfstring{RT-$\Lambda$CC}{RT-LambdaCC} propagation}

We partition the neutral and ionized cluster operators into non-core and core-sector components,
\begin{equation}
T^{(N)}=T_r+T_c,
\qquad
S(t)=S_r(t)+S_c(t).
\label{eq:partition}
\end{equation}
Because $T_c|\phi_c^{(N-1)}\rangle=0$, the appropriate real-time initial condition is
\begin{equation}
S_r(0)=T_r,
\qquad
S_c(0)=0,
\label{eq:shift_initial}
\end{equation}
or, equivalently,
\begin{equation}
S_r(t)=T_r+\Delta S_r(t),
\qquad
\Delta S_r(0)=0.
\label{eq:shifted_amplitudes}
\end{equation}
The propagated amplitudes $\Delta S_r(t)$ describe the additional relaxation and screening induced by the core hole on top of the neutral-state correlation already present at $t=0$.

We interpret this shifted initialization as the leading non-core contribution of the dCC construction. Within a single-reference framework, the dCC ansatz recovers the exact Green's-function limit as its neutral- and ionized-sector cluster expansions become complete~\cite{peng2024exploring}. In the Baker--Campbell--Hausdorff (BCH) expansion of the dCC product of these exponentials, the non-core operators combine directly, whereas the remaining commutators contain at least one core-sector operator and describe higher-order hole-mediated couplings. RT-$\Lambda$CC retains the leading correlated projection through Eq.~\eqref{eq:shift_initial} while leaving those higher-order core-sector commutators out of the propagated equations of motion. Section~S2.4 of the Supporting Information details the connection to the dCC hierarchy.

\subsection{\texorpdfstring{$\Lambda$-weighted}{Lambda-weighted} overlap and spectral decomposition}

We write the RT-$\Lambda$CC core-hole Green's function as
\begin{equation}
G_c^{R}(t)
\simeq
-i\Theta(t)e^{-iE_{\mathrm{CC}}^{(N)}t}
N_c(t)\,
O_c^{\Lambda}(t),
\label{eq:rtlambda_green}
\end{equation}
where $N_c(t)$ is the scalar phase/cumulant factor generated by the shifted real-time propagation and
\begin{equation}
O_c^{\Lambda}(t)
=
\left\langle\phi_0^{(N)}\right|
(1+\Lambda^{(N)})e^{-T^{(N)}}
a_c^\dagger e^{S(t)}
\left|\phi_c^{(N-1)}\right\rangle
\label{eq:lambda_overlap_full}
\end{equation}
projects the evolving core-hole state onto the correlated neutral coupled-cluster left state. Using Eq.~\eqref{eq:shifted_amplitudes}, the static non-core correlation cancels from the overlap, yielding
\begin{equation}
O_c^{\Lambda}(t)
=
\left\langle\phi_0^{(N)}\right|
(1+\Lambda^{(N)})e^{-T_c}
e^{\Delta S_r(t)}
\left|\phi_0^{(N)}\right\rangle .
\label{eq:lambda_overlap_compact}
\end{equation}
Section~S2.2 of the Supporting Information details the step-by-step
cancellation leading to Eq.~\eqref{eq:lambda_overlap_compact}.
At the coupled-cluster singles-and-doubles (CCSD) level,
\begin{equation}
O_c^{\Lambda}(t)
=
1+O_2(t)+O_{3a}+O_{3b}(t),
\label{eq:overlap_partition}
\end{equation}
Table~\ref{tab:overlap_summary} summarizes the three terms. $O_2(t)$ strongly
dominates the numerical hierarchy.
Section~S2.3 of the Supporting Information provides explicit contractions and the corresponding time- and frequency-domain comparisons. The
$A$-local, $B$-local, and CT labels used below resolve orbital terms within
the $O_2(t)$ sector. Consequently, we do not identify them with separate $O_2$ and $O_3$
mechanisms.

\begin{table}[t]
  \centering
  \caption{Compact summary of the $\Lambda$-weighted overlap terms at the
  CCSD level.}
  \label{tab:overlap_summary}
  \setlength{\tabcolsep}{3pt}
  \renewcommand{\arraystretch}{1.14}
  \begin{tabular}{@{}l p{0.73\columnwidth}@{}}
    \toprule
    \textbf{Term} & \textbf{Role} \\
    \midrule
    $O_2(t)$ & non-core left/right overlap response \\
    $O_{3a}$ & static core-sector correction \\
    $O_{3b}(t)$ & time-dependent core-mediated contribution \\
    \bottomrule
  \end{tabular}
\end{table}

The same overlap decomposition lets us classify the components underlying the many-body satellite response as acceptor-local, donor-local, or intermolecular charge-transfer-like. We propagate in the canonical molecular-orbital basis and apply the fragment-resolved analysis through a unitary post-processing transformation, which leaves the total overlap unchanged. Section~S4 and Figure~S1 of the Supporting Information document the fragment-canonical construction and classification rules.

\subsection{Scope of the approximation}

RT-$\Lambda$CC therefore occupies an intermediate level between conventional single-exponential RT-EOM-CC and the full dCC construction. It retains the correlated sudden-ionization initial condition, the neutral-state $\Lambda$ projection required for the spectral overlap, and the single-exponential real-time propagation. Relative to dCC, it omits higher-order core-sector BCH commutators from the propagated Hamiltonian. The benchmark calculations below quantify how much of the dCC response this reduced construction retains before we use it to analyze nonlocal screening in the water dimer.
Section~S2 of the Supporting Information collects the supporting derivations for the RT-$\Lambda$CC formulation, and Section~S2.4 treats the retained and omitted dCC terms explicitly.

\section{Results and Discussion}

Section~S1 of the Supporting Information summarizes the molecular geometries,
propagation and output time steps, propagation intervals, fragment integration,
and analysis conventions.
Here ``ultrafast'' denotes the $\sim$10~fs fixed-nuclei electronic response.

\subsection{Correlated sudden-state propagation and relation to RT-dCC}

\begin{figure*}[t]
  \centering
  \includegraphics[width=0.42\textwidth]{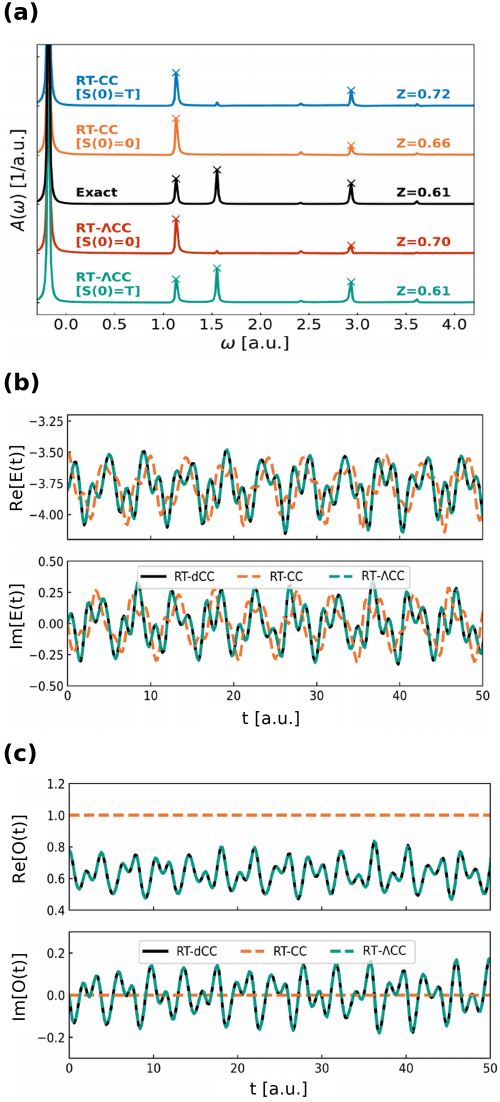}
  \caption{Compact validation of the two ingredients in RT-$\Lambda$CC.
  (a) Three-site single-impurity Anderson-model spectral functions. The black
  curve is the exact reference obtained by direct diagonalization, which
  RT-dCC also recovers for this system. The other
  curves show the indicated RT-CC/RT-$\Lambda$CC initial-condition and overlap
  combinations. Panels (b) and (c) show the real and imaginary parts of the instantaneous core-hole energy
  $E(t)$ and overlap $O(t)$, respectively, for conventional RT-CC, correlated shifted-start
  RT-$\Lambda$CC, and RT-dCC.}
  \label{fig:method_validation}
\end{figure*}

The three-site single-impurity Anderson model provides a controlled benchmark for separating the roles of the correlated initial condition and the $\Lambda$-weighted overlap. As seen from Figure~\ref{fig:method_validation}a, shifting the initial condition alone does not recover the exact quasiparticle (QP) weight, or main-pole residue: conventional real-time coupled-cluster (RT-CC) with $S(0)=T$ gives $Z=0.72$, compared with the exact value $Z=0.61$. Likewise, including the $\Lambda$ overlap while retaining $S(0)=0$ yields $Z=0.70$ and does not reproduce the correct satellite intensities. Only the combination of the correlated sudden-state initialization and the $\Lambda$-weighted overlap recovers $Z=0.61$ together with the satellite positions available within the CCSD excitation manifold. The two ingredients therefore play complementary roles: the correlated shifted-start initialization places the propagation on the correlated sudden-ionization manifold, whereas the overlap restores the corresponding projection onto the neutral coupled-cluster left state.

A direct comparison with real-time dCC (RT-dCC) shows which features the lower-complexity construction retains (Figure~\ref{fig:method_validation} and Section~S3 of the Supporting Information). As seen from Figures~\ref{fig:method_validation}b and~\ref{fig:method_validation}c, the correlated shifted-start RT-$\Lambda$CC dynamics closely follow RT-dCC for both the real and imaginary parts of the instantaneous core-hole energy $E(t)$ and the corresponding $\Lambda$-weighted overlap function $O(t)$. The algebra in Section~S3 shows that single-exponential RT-$\Lambda$CC omits the core-mediated BCH couplings present in RT-dCC, whereas conventional RT-CC fixes the associated overlap contribution to unity and misses its time dependence~\cite{abraham2026elucidating}. For this benchmark, RT-$\Lambda$CC reproduces the principal dCC-like real-time behavior associated with the correlated neutral sector. 

\subsection{Nonlocal valence screening at the equilibrium hydrogen bond}

We next consider acceptor-site O~1s ionization of the equilibrium water dimer using the high-level coupled-cluster singles and doubles with perturbative triples [CCSD(T)] structure from Klopper \emph{et al.}~\cite{klopper2000waterdimer}. This well-characterized donor--acceptor minimum provides a common high-level origin for the prescribed O--H and O$\cdots$O scans and avoids folding a basis-dependent reoptimization into the geometry trends. Earlier studies established that donor and acceptor sites are spectroscopically distinct, that ionized water dimers can couple strongly to proton-transfer coordinates, and that ionization energies and aggregate satellite structure evolve with geometry~\cite{vila2022water_dimer_xps,felicissimo2005water_dimer_core,segarra-marti2012photophysics,bjorneholm1999between,bodi2014protonation}. The advance sought here is different: we resolve how a site-local core hole distributes correlated satellite weight among acceptor-local, donor-local, and interfragment configurations, establish the intermolecular origin of those pathways by separating the monomers, and identify the donor-orbital amplitude groups that carry their ultrafast modulation.

Acceptor-site ionization provides a clean window onto the prompt electronic response across an intact hydrogen bond. Because the donor-ionized surface has the stronger displacement along proton-transfer coordinates~\cite{felicissimo2005water_dimer_core}, placing the core hole on the acceptor reduces the competing nuclear driving force. The subsequent O$\cdots$O and donor O--H scans then separate the existence of intermolecular screening from its geometry-dependent redistribution.

The RT-$\Lambda$CCSD O~1s photoelectron spectrum separates naturally into three energy regimes when we analyze the overlap components in the fragment-canonical orbital basis (Figure~\ref{fig:equilibrium_overview}). Below approximately 16~eV relative to the main quasiparticle peak, valence excitations localized on the neighboring donor molecule dominate the response. Local shake-up excitations on the core-ionized acceptor dominate the intermediate region, approximately 18--25~eV. At higher energy, above approximately 26~eV, intermolecular charge-transfer-like configurations acquire increasing weight.

Representative donor-local motifs include B:$3a_1$ excitations at relative
energies of 10.683, 12.820, 14.957, and 15.811~eV. Acceptor-local motifs
dominate prominent features from approximately 18 to 25~eV, whereas
interfragment excitations appear among the higher-energy features.
Table~\ref{tab:h2odimer_satellites_631ppgss} reports the full peak-level
assignments. As documented by the direct comparison in Section~S8 and
Table~S4 of the Supporting Information, the acceptor-local assignments are
also consistent with water-monomer SAC-CI shake-up
energies~\cite{Sankari2006SACCI}. The monomer comparison therefore anchors the
local baseline. The donor-local and interfragment components identified below
are dimer-specific contributions that have no monomer counterpart.

\begin{table}[!t]
\centering
\caption{Principal satellite features of the equilibrium water dimer that we
calculated with the 6-31++G** basis. The main quasiparticle peak lies at
$-543.562$~eV, and $\Delta E$ is the satellite separation from that peak. The
motif weight is the fraction of the ranked component area for the listed
satellite transition. The motif families identify the dominant ranked
components rather than a complete state-by-state decomposition.}
\label{tab:h2odimer_satellites_631ppgss}
\setlength{\tabcolsep}{3pt}
\renewcommand{\arraystretch}{1.08}
\begin{tabular}{@{}c c p{0.46\columnwidth}@{}}
\toprule
\textbf{$\Delta E$ (eV)} & \textbf{Motif weight (\%)} & \textbf{Dominant motif family} \\
\midrule
10.683 & 40 & $B{:}1b_1 \rightarrow B{:}5a_1$ and $B{:}3a_1 \rightarrow B{:}2b_1$ \\
12.820 & 54 & $B{:}3a_1 \rightarrow B{:}4b_2$ and $A{:}1b_1 \rightarrow A{:}4a_1$ \\
14.102 & 79 & $B{:}1b_1 \rightarrow B{:}5a_1$ \\
14.957 & 77 & $B{:}3a_1 \rightarrow B{:}7a_1$ \\
15.811 & 79 & $B{:}3a_1 \rightarrow B{:}2b_1$ \\
17.948 & 43 & $A{:}3a_1 \rightarrow A{:}4a_1$ \\
21.366 & 58 & $A{:}1b_1 \rightarrow A{:}5a_1$ \\
22.221 & 31 & $A{:}1b_1 \rightarrow A{:}5a_1$ \\
25.212 & 15 & $A{:}1b_2,A{:}1b_2 \rightarrow A{:}4b_2,A{:}4b_2$ \\
25.640 & 13 & $A{:}1b_2,A{:}3a_1 \rightarrow A{:}4b_2,A{:}5a_1$ \\
26.922 & 28 & $A{:}1b_1 \rightarrow B{:}2b_2$ \\
27.349 & 41 & $A{:}3a_1 \rightarrow A{:}7a_1$ \\
28.631 & 22 & $A{:}3a_1 \rightarrow A{:}7a_1$ \\
29.486 & 41 & $A{:}3a_1 \rightarrow B{:}7a_1$ \\
30.768 & 24 & $A{:}1b_2,A{:}1b_2 \rightarrow A{:}4b_2,A{:}6a_1$ \\
\bottomrule
\end{tabular}
\end{table}

More importantly, the dimer-specific contributions show that the response to a
localized acceptor core hole extends onto the neighboring donor. Within
the fragment-resolved component analysis, integration over the 540--575~eV
binding-energy range gives normalized classified component-area fractions of
77.5\% acceptor-local, 2.7\% donor-local, and 19.8\%
charge-transfer-like character at equilibrium. Thus, 22.5\% of the classified
satellite component area involves donor-local or CT-like channels.

\begin{figure*}[t]
  \centering
  \includegraphics[width=\textwidth]{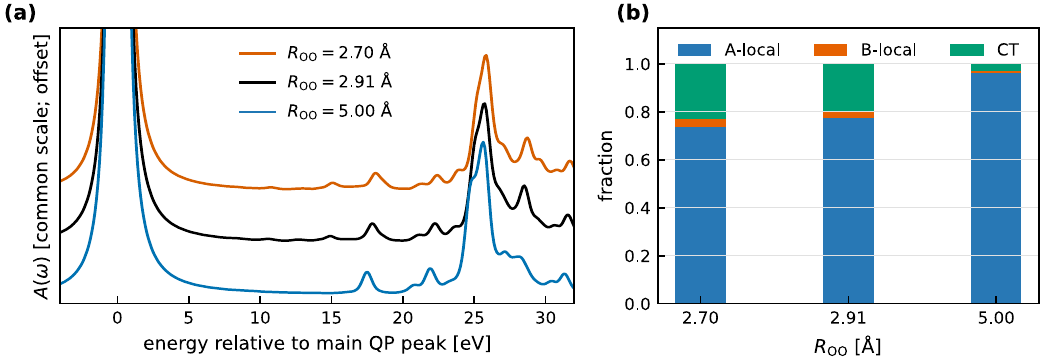}
  \caption{Intermolecular-separation control at fixed intramolecular
  geometries and relative orientation. (a) RT-$\Lambda$CCSD O~1s spectra for
  $R_{\mathrm{OO}}=2.70$, 2.91, and 5.00~\AA{} across the full analyzed
  satellite manifold. We align the quasiparticle peaks at zero energy and
  use one common intensity scale while shifting the spectra vertically for
  clarity. We generate the curves from the archived time responses with
  damping parameter $\eta=0.01$~a.u. (b) Stacked
  bar chart of the $A$-local, $B$-local, and CT fractions. We integrate them
  over the full 540--575~eV analyzed binding-energy range and normalize them
  after excluding the O~1s quasiparticle contribution.}
  \label{fig:separation_control}
\end{figure*}

To isolate the intermolecular origin of this response, we varied the O$\cdots$O separation while keeping both intramolecular geometries and the relative orientation fixed (Figure~\ref{fig:separation_control}, Section~S6, and Table~S2 of the Supporting Information). At the equilibrium separation, $R_{\mathrm{OO}}=2.91$~\AA{}, the $A$-local, $B$-local, and CT fractions are 0.775, 0.027, and 0.198, respectively. When we separate the monomers to $R_{\mathrm{OO}}=5.00$~\AA{}, the $B$-local and CT fractions decrease to 0.006 and 0.031, while the $A$-local fraction rises to 0.962. The same separation suppresses the low-energy donor-local satellite features. By contrast, shortening the intermolecular separation to 2.70~\AA{} produces only a modest redistribution relative to equilibrium. Because we keep the donor O--H geometry unchanged in this control, the collapse of the nonlocal component at 5.00~\AA{} supports intermolecular coupling as its origin.

To determine which many-electron pathways carry the nonlocal response in time,
we combine the explicit orbital indexing of the CC amplitudes with a short-time
Fourier transform (STFT). The integrated overlap decomposition establishes how
much satellite weight belongs to each fragment class. The orbital-resolved
quantity below then identifies which occupied-orbital-indexed amplitude groups
carry that response and how their modulation evolves. For an occupied
fragment-canonical orbital $p$, we define the excitation weight
\begin{equation}
H_p(t)=
\sum_{\mu:\,p\in h(\mu)}
w_{\mu p}\left|\Delta s_\mu(t)\right|^2,
\label{eq:orbital_weight}
\end{equation}
where $\mu$ labels a propagated CCSD excitation amplitude, $h(\mu)$ is its set
of occupied indices, and $w_{\mu p}$ accounts for the index occurrence in the
spin-resolved CCSD representation. Section~S5 of the Supporting Information
gives the corresponding expanded expression. Because the amplitudes are
dynamical coordinates of the propagated coupled-cluster state, their squared
magnitudes define a positive relative weight for the overlapping amplitude
group carrying occupied index $p$. Subtracting the inherited neutral-state
amplitudes gives $H_p(0)=0$.

$H_p(t)$ connects the propagated CC wavepacket to a chemically indexed set of
excitation coordinates: a larger value means that changes in amplitudes
containing $p$ carry more weight in the right-hand CC parametrization. Its
representation dependence provides useful resolution when the representation
is chosen deliberately. In the fragment-canonical basis used here, $p$ retains
both orbital and fragment identity, allowing the donor and acceptor amplitude
groups to be followed separately. A one-particle density instead contracts the
left and right states into an observable charge distribution. In contrast,
$H_p(t)$ retains the identity of the correlated excitation-amplitude groups
that produce that distribution. The two analyses therefore answer different questions: density
describes where charge resides, whereas $H_p(t)$ identifies which
orbital-indexed many-electron coordinates carry the evolving response. Because
a double excitation can contain more than one occupied index, an amplitude can
contribute to multiple $H_p(t)$ groups. We therefore interpret these groups as
overlapping orbital-indexed measures of the evolving response.

We obtain the time--frequency content from the STFT magnitude
\begin{equation}
A_p(\Omega,\tau)=
\left|
\int dt\,W(t-\tau)H_p(t)e^{i\Omega t}
\right|,
\label{eq:weight_stft}
\end{equation}
where $W(t-\tau)$ is a finite window centered at $\tau$. The magnitude
$A_p(\Omega,\tau)$ reports the modulation carried by group $p$ near time
$\tau$, and its square $F_p(\Omega,\tau)=A_p(\Omega,\tau)^2$ is the
corresponding spectrogram power. When reporting $\Omega$ on an energy axis, we
denote it by $E_{\mathrm{mod}}$ and write the same magnitude as
$A_p(E_{\mathrm{mod}},\tau)$. This local frequency decomposition captures
both oscillatory structure and non-exponential changes in the propagated
amplitudes without imposing a single decay model.

The sudden-ionization wavepacket connects the STFT to the photoelectron
satellite manifold. If
$I_K=E_K^{N-1}-E_0^N$ and $I_L=E_L^{N-1}-E_0^N$ are the binding energies of two
ionized-state components, a bilinear contribution can oscillate at their
difference frequency. Expressed in energy units, this gives
\begin{equation}
E_{\mathrm{mod}}^{KL}=|I_K-I_L|.
\label{eq:modulation_difference}
\end{equation}
When $L$ is the QP state,
$E_{\mathrm{mod}}^{K,\mathrm{QP}}=I_K-I_{\mathrm{QP}}\equiv\Delta E_K$ is the
satellite's binding energy relative to the quasiparticle peak. QP--satellite
beating therefore provides a direct bridge between an orbital-indexed STFT band
and the corresponding relative satellite-energy region. Satellite--satellite
pairs and nonlinear amplitude couplings can contribute additional difference
frequencies, so we use this bridge to identify coherent energy regions rather
than to invert each STFT maximum into a unique stationary state. Fully
state-resolved assignments would require biorthogonal EOM-CC or response-state
projections~\cite{pedersen2021interpretation}.

Short-time Fourier analysis has previously retained temporal information in
field-induced polarization responses from fixed-nuclei dimers~\cite{akama2010stft}.
Applying it here to orbital-indexed CC excitation weights extends that idea to
the correlated core-ionized wavepacket: the maps resolve which
fragment-canonical occupied-orbital groups carry the screening dynamics and
how their characteristic modulation energies change in time.

For independent regeneration, we fit and subtract a quartic polynomial from
each archived $H_p(t)$ trace and apply a symmetric 827-sample Hann window to
the 0.1~a.u. output grid. The window spans 1.998~fs, its center advances by
103 samples (0.249~fs), and fourfold zero padding gives a 3308-point transform.
We first calculate $A_p=|\widetilde H_p|$ directly and then divide every map in
a figure by one common reference $A_{\mathrm{ref}}$, the 99.5th percentile over
the displayed panels. We cap the normalized color scale at 1. This linear,
figure-wide rescaling preserves cross-panel amplitude ratios below that ceiling
through one common normalization across all panels. As in earlier real-time STFT
analysis~\cite{akama2010stft}, the approximately 2~fs window gives the nominal
Fourier energy scale $h/T_{\mathrm{win}}\approx 2.1$~eV. The smaller
window-center increment provides finer temporal sampling, while the
approximately 2~fs window sets the intrinsic energy resolution. The resulting
energy resolution is well matched to comparisons between the separated 10--16
and 20--30~eV response regions.

\begin{figure*}[t]
  \centering
  \includegraphics[width=0.98\textwidth]{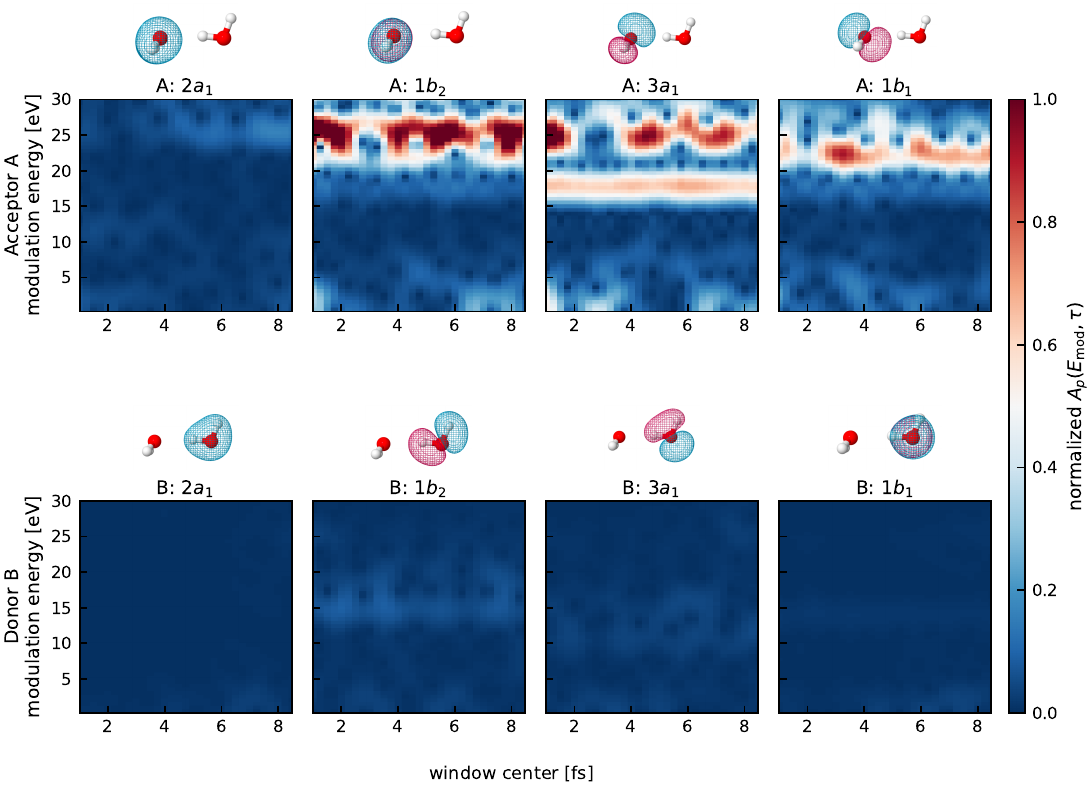}
  \caption{Equilibrium-geometry STFT magnitudes $A_p(E_{\rm mod},\tau)$ of the orbital-resolved CC
  excitation weights for the acceptor ($A$) and donor ($B$) valence
  orbitals. The molecular-orbital isosurface above each map identifies the corresponding
  occupied fragment-canonical orbital. The horizontal coordinate
  is the STFT window center and the vertical coordinate is the local modulation
  energy $E_{\mathrm{mod}}$. Dark red marks values near the upper end
  of the common normalized $A_p(E_{\mathrm{mod}},\tau)$ scale and therefore identifies the largest
  local modulation magnitude among the displayed orbital-indexed groups at
  fixed time and energy. The donor B:$3a_1$
  group provides the equilibrium reference for the O--H scan in
  Figure~\ref{fig:proton_control}. }
  \label{fig:equilibrium_stft_main}
\end{figure*}

The equilibrium $A_p(E_{\mathrm{mod}},\tau)$ magnitudes in
Figure~\ref{fig:equilibrium_stft_main} establish the cross-orbital reference
before we displace the donor O--H coordinate. At fixed
$(E_{\mathrm{mod}},\tau)$, the common $A_p/A_{\mathrm{ref}}$ scale preserves
the amplitude differences among all eight displayed groups. Within each panel,
the same scale shows how a given group's modulation evolves with time and
energy. The common analysis therefore preserves the physically weaker donor
response alongside the stronger acceptor response.

At equilibrium, the dark-red band of A:$1b_2$ persists near 24--27~eV over
most of the displayed approximately 1.0--8.5~fs interval, while A:$3a_1$ and A:$1b_1$ show
stronger, time-dependent patches in the approximately 21--26~eV region.
A:$3a_1$ also carries a persistent lower band near 18~eV. On the same scale,
the donor panels remain much weaker: B:$1b_2$ is most evident near 15~eV and
B:$3a_1$ across approximately 10--16~eV. These horizontal bands show that the
dominant modulation-energy regions persist during the ultrafast electronic
response, whereas changes from red to lighter colors show that their local
Fourier amplitudes vary with the window center. At the approximately 2~fs
window resolution, the color evolution locates when each orbital-indexed group
builds up or weakens. The pattern therefore identifies the amplitude groups
that carry the strongest correlated-wavepacket modulation and the time
intervals over which they are enhanced.

The donor interval contains QP-relative satellite energies of 10.683, 12.820,
14.957, and 15.811~eV whose ranked motifs involve B:$3a_1$. Likewise, the
A:$1b_1$ and A:$1b_2$ bands lie near the 21.366/22.221 and
25.212/25.640~eV satellite separations carrying those orbital labels. This
orbital-specific alignment connects the acceptor- and donor-indexed dynamics to
their respective satellite manifolds. The nominal $\sim2.1$~eV resolution
separates these broad manifolds and supports orbital-specific comparison with
their satellite regions.

\subsection{Proton displacement redistributes local and intermolecular screening}

\begin{figure*}[!t]
  \centering
  \includegraphics[width=0.98\textwidth]{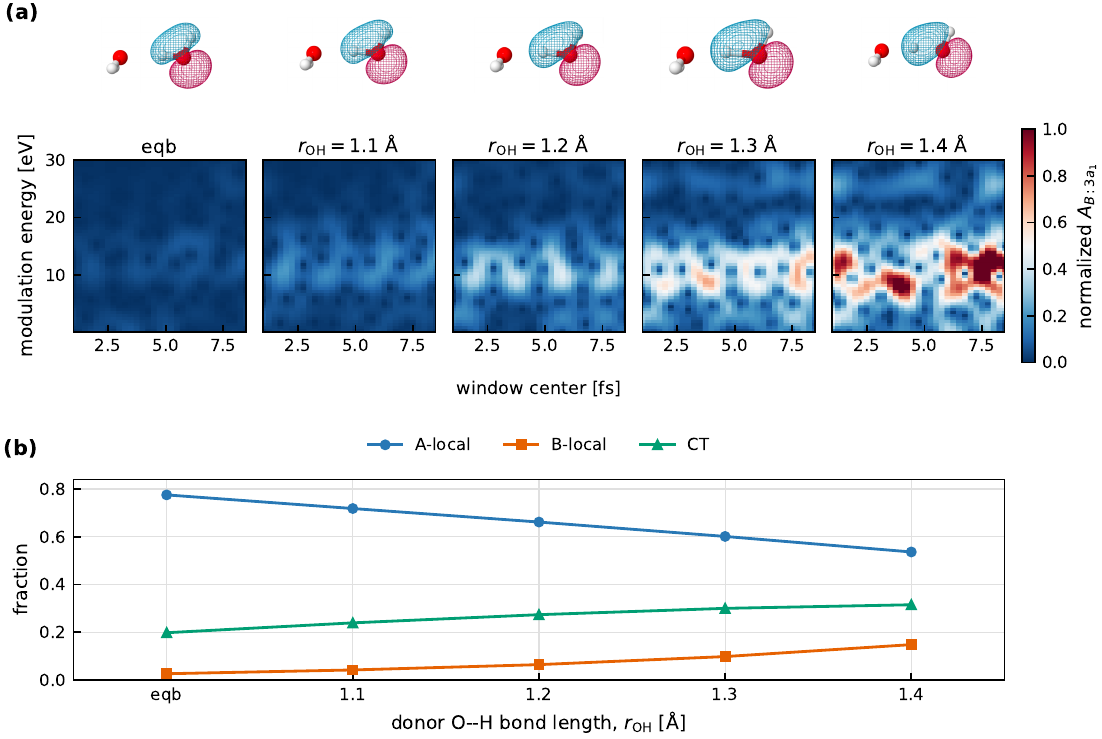}
  \caption{Donor O--H displacement redistributes the fragment-resolved
  response. (a) Direct STFT magnitude
  $A_p(E_{\mathrm{mod}},\tau)$ of the
  orbital-resolved CC excitation weight for the donor-localized B:$3a_1$
  orbital after division by one five-geometry reference,
  $A_{\mathrm{ref}}$, defined as the 99.5th percentile over all five maps.
  The molecular-orbital isosurface above each map shows the
  corresponding B:$3a_1$ orbital at that geometry. Dark red marks values near the upper end
  of the common normalized scale and therefore identifies the
  window centers and modulation energies at which the B:$3a_1$-indexed
  amplitude group modulates most strongly. All five panels share a 0.999--8.473~fs
  window-center axis and color scale. 
  (b) Fractions of the integrated satellite component area classified as $A$-local,
  $B$-local, and CT. We normalize them by the total classified satellite component area at each geometry after excluding the
  O~1s quasiparticle contribution.
  In both panels, ``eqb'' denotes the equilibrium geometry, corresponding to
  the donor O--H distance $r_{\mathrm{OH}}=0.964$~\AA{}.}
  \label{fig:proton_control}
\end{figure*}

We next ask how the hydrogen-bonded proton coordinate reorganizes the
core-hole response. In a fixed-nuclei scan (Figure~\ref{fig:proton_control}),
we elongate the donor O--H bond from its equilibrium value of 0.964~\AA{} to
1.1, 1.2, 1.3, and 1.4~\AA{}, while retaining the intermolecular framework.
For each geometry, the panels in Figure~\ref{fig:proton_control}a follow the
donor B:$3a_1$-indexed dynamics, and the dots in
Figure~\ref{fig:proton_control}b measure the integrated redistribution among
acceptor-local, donor-local, and charge-transfer-like satellite components.

The common scale in Figure~\ref{fig:proton_control}(a) reveals a systematic
activation of the donor B:$3a_1$ group. Its 10--16~eV band is weak at
equilibrium and 1.1~\AA{}, strengthens at 1.2~\AA{}, becomes repeatedly
prominent at 1.3~\AA{}, and reaches the dark-red end of the scale at
1.4~\AA{}. At the longest bond length, the dominant weight moves between the
approximately 8--10 and 11--13~eV regions: it is strongest near 11--13~eV
early in the response, shifts toward 8--10~eV around 3--5~fs, and broadens
across 9--13~eV after approximately 6~fs. At the approximately 2~fs window
resolution, this evolution shows that O--H elongation changes both the strength
of the donor-indexed response and the mixture of modulation energies carried by
that group.

A QP-referenced coherence analysis anchors this time--frequency evolution to
the core-hole wavepacket: the principal B:$3a_1$ modulation maxima covary with
features in the QP-referenced core-hole response along the O--H scan.
Definitions, trace lengths, coherence values, detrending tests, and numerical
peak locations are reported in Section~S5, Figure~S2, and Table~S1 of the
Supporting Information. Together with the 10--16~eV energy-region alignment in
Figure~\ref{fig:equilibrium_stft_main}, this covariance connects the growing
B:$3a_1$ amplitude modulation to the donor-local satellite manifold.

Figure~\ref{fig:proton_control}(b) shows the corresponding redistribution of
integrated satellite weight. The acceptor-local fraction decreases
monotonically from 0.775 at equilibrium to 0.536 at
$r_{\mathrm{OH}}=1.4$~\AA{}. Over the same coordinate range, the donor-local
fraction increases from 0.027 to 0.148, while the intermolecular
charge-transfer-like fraction increases from 0.198 to 0.315. The combined
$B$-local+CT fraction therefore rises from 0.225 to approximately 0.464, and
the donor-local contribution exhibits the largest relative change, increasing
by approximately a factor of 5.5.

Because the nuclei remain fixed during each propagation, each calculation
measures the electronic response at a prescribed proton coordinate. That
separation is useful here: it shows
that proton displacement reshapes the available screening pathways even before
nuclear motion is propagated, redistributing satellite weight from
acceptor-local relaxation toward donor-local and interfragment channels.

A targeted basis-robustness test gives the same qualitative redistribution
with aug-cc-pVDZ (Section~S7, Figure~S3, and Table~S3 of the Supporting
Information). At equilibrium, the aug-cc-pVDZ fractions are 0.711, 0.043, and
0.246 for $A$-local, $B$-local, and CT character, respectively. At
$r_{\mathrm{OH}}=1.4$~\AA{}, they become 0.451, 0.197, and 0.352. Although
the absolute fractions are basis dependent, both aug-cc-pVDZ and 6-31++G**
show a substantial loss of acceptor-local character and increases in both
donor-local and CT character upon O--H elongation.

The two panels supply distinct pieces of the mechanism. Panel (b) quantifies
how much screening weight moves into donor-involving channels, whereas panel
(a) identifies B:$3a_1$ as a donor-indexed amplitude group that carries this
change and resolves its evolution in time and modulation energy. Their joint
trend links the geometry-enhanced donor-local satellite weight to a stronger
donor-orbital contribution in the correlated wavepacket. In larger
hydrogen-bonded networks or pump--probe simulations, the same orbital-resolved
analysis can identify when and through which fragment-canonical orbitals a
screening pathway develops, even when transient spectral features overlap.

\section*{Conclusions and outlook}

{
An acceptor-localized O~1s core hole in the water dimer produces a finite
donor-involving screening component. At equilibrium, 22.5\% of the normalized
fragment-classified satellite component area is donor-local or CT-like, even
though acceptor-local shake-up remains dominant. Increasing
$R_{\mathrm{OO}}$ to 5.00~\AA{} at fixed intramolecular geometries reduces the
donor-local and CT fractions to 0.6\% and 3.1\%, respectively. Their collapse
upon separation supports intermolecular coupling as the origin of the
equilibrium nonlocal component.

The fixed-nuclei O--H scan shows that the proton coordinate strongly
redistributes these channels. Elongating the donor O--H bond from 0.964 to
1.4~\AA{} increases the combined donor-local and CT-like fraction from 22.5\%
to 46.4\%, while the donor-local fraction rises from 2.7\% to 14.8\%. The same
qualitative trend with 6-31++G** and aug-cc-pVDZ provides a basis-robustness
test without implying basis-set convergence.

Along the same scan, the common-scale STFT shows that the donor B:$3a_1$
excitation-weight modulation becomes progressively stronger and changes its
time--energy distribution. Its alignment with the donor-local satellite region
and covariance with the QP-referenced core-hole response connect this
orbital-indexed dynamics to the growing donor-involving screening channels.
The STFT therefore adds information absent from the integrated spectrum: it
identifies which occupied-orbital group carries the response and when that
group is activated.

RT-$\Lambda$CC enables the connection between spectrum and mechanism. Unlike
a peak-only analysis, the framework determines the quasiparticle--satellite
balance and resolves the satellite response into chemically identifiable
fragment pathways. Correlated sudden-state initialization incorporates neutral-state correlation,
the $\Lambda$-weighted overlap supplies the spectral projection and the
fragment-resolved decomposition, and single-exponential propagation retains
the practical structure of RT-EOM-CC while following the leading behavior of
the dCC reference. Together, these ingredients establish a route from a
site-local core-hole perturbation to chemically resolved many-electron
screening pathways. The water dimer provides the minimal controlled benchmark
across one hydrogen bond, laying the foundation for extending this framework
to larger hydrogen-bonded systems and their X-ray spectroscopy.
}

\section*{Author contributions}
Vibin Abraham: Data curation, Formal analysis, Investigation, Methodology,
Software, Validation, Visualization, Writing -- original draft. Bo Peng:
Conceptualization, Formal analysis, Funding acquisition, Investigation,
Methodology, Project administration, Resources, Software, Supervision,
Validation, Visualization, Writing -- original draft, Writing -- review \&
editing.

\section*{Conflicts of interest}
There are no conflicts to declare.

\section*{Data availability}
The Zenodo dataset ``Real-time coupled-cluster water-dimer core-hole dynamics
data'' provides the data supporting this article at
\url{https://doi.org/10.5281/zenodo.21939925}. The archive contains the
paper-level complex Green's-function signals and overlap decompositions,
orbital-resolved CC excitation-weight and short-time Fourier-transform arrays, component-area
summary tables, metadata, and plotting scripts. The Supporting Information
provides additional theoretical derivations, analysis definitions, and
numerical validation.

\section*{Acknowledgements}
This work was supported by the Early Career Research Program of the U.S. Department of Energy, Office of Science, under Grant No.~FWP 83466. V.A. acknowledges Niri Govind for helpful discussions.

% ============================================================

\clearpage
\onecolumngrid
\begin{center}
  {\LARGE\bfseries Supporting Information\par}
  \vspace{0.5em}
  {\large\bfseries Geometry-Dependent Nonlocal Valence Screening Following
  Core Ionization of the Water Dimer\par}
  \vspace{0.5em}
  {\normalsize Vibin Abraham and Bo Peng\par}
\end{center}
\vspace{1em}
\twocolumngrid

% Preserve the standalone SI numbering within the combined arXiv document.
\setcounter{section}{0}
\setcounter{subsection}{0}
\setcounter{equation}{0}
\setcounter{figure}{0}
\setcounter{table}{0}
\renewcommand{\thesection}{S\arabic{section}}
\renewcommand{\thesubsection}{\thesection.\arabic{subsection}}
\renewcommand{\theequation}{S\arabic{equation}}
\renewcommand{\thefigure}{S\arabic{figure}}
\renewcommand{\thetable}{S\arabic{table}}
\renewcommand{\theHsection}{S.\arabic{section}}
\renewcommand{\theHsubsection}{S.\arabic{section}.\arabic{subsection}}
\renewcommand{\theHequation}{S.\arabic{equation}}
\renewcommand{\theHfigure}{S.\arabic{figure}}
\renewcommand{\theHtable}{S.\arabic{table}}

\section{Computational and Analysis Protocol}
\label{sec:si_protocol}
% ============================================================

We use the equilibrium donor--acceptor water-dimer geometry that Klopper
\emph{et al.} optimized with coupled-cluster (CC) singles and doubles (CCSD)
plus perturbative triples [CCSD(T)]~\cite{klopper2000waterdimer}.
Molecule $A$ is the hydrogen-bond acceptor and carries the suddenly created
O~$1s$ hole. Molecule $B$ is the hydrogen-bond donor. The primary calculations
use correlated shifted-start real-time $\Lambda$-coupled-cluster theory
(RT-$\Lambda$CC) at the singles-and-doubles level (RT-$\Lambda$CCSD) with the
6-31++G** basis.
Endpoint calculations with aug-cc-pVDZ at the equilibrium geometry and at
$r_{\mathrm{OH}}=1.4~\text{\AA}$ test the basis robustness of the spectral
redistribution. Because we compare only two geometries and two basis sets, we
do not present these calculations as a basis-set convergence study.

We propagated the core-ionized water-dimer wavefunction for 400~a.u.
(9.68~fs) following sudden removal of the acceptor O~$1s$ electron at every
fixed geometry. We used an internal propagation step of 0.05~a.u. (1.21~as) and
recorded the response every 0.1~a.u. (2.42~as). The paper-level Green's
functions and overlap decompositions retain this full interval. The compact
orbital-resolved CC excitation-weight exports used for the B:$3a_1$
short-time Fourier transform (STFT) analysis extend to 9.499~fs at all five
geometries and yield window centers from 0.999 to 8.473~fs, as
Section~\ref{sec:si_orbital_weight} documents. In the donor O--H scan, we
elongated the hydrogen-bonded donor bond from its equilibrium value of
0.964~\text{\AA} to 1.1, 1.2, 1.3, and 1.4~\text{\AA} while holding the
remaining coordinates fixed. In the intermolecular control, we set
$R_{\mathrm{OO}}$ to 2.70, 2.91, and 5.00~\text{\AA} while holding both
intramolecular geometries and the relative orientation of the monomers fixed.
These two scans therefore probe
different coordinates: the O--H scan changes the donor intramolecular
geometry within the dimer, whereas the O$\cdots$O scan changes only the
intermolecular separation.

For the finite-record spectra in Figure~3(a) of the main article and
Figure~\ref{fig:basis-comparison-eqb-r14}, we multiply the exponentially
damped Green's-function signal by a terminal raised-cosine taper over the
final 25\% of the time record before Fourier transformation. This processing
suppresses truncation sidelobes but does not extend the propagation or improve
the intrinsic spectral resolution. A sensitivity check against the untapered
spectra leaves the principal satellite centers unchanged on the 0.107~eV
displayed grid, changes their widths by at most 0.02~eV, and changes the
integrated satellite weight by less than 0.004\%.

We obtained fragment fractions by integrating the absolute
fragment-classified component area over the 540--575~eV analyzed
binding-energy range, excluding the O~$1s$ quasiparticle contribution, and
normalizing by the total classified satellite component area at the same
geometry. The resulting $A$-local, $B$-local, and charge-transfer-like (CT)
fractions are component-analysis measures in the fragment-canonical orbital
representation. Consequently, they are not fragment charges or probabilities.

For the three-site single-impurity Anderson-model benchmark, the Hamiltonian
contains one correlated impurity and two bath sites with $U=2.0$~a.u.,
$\mu_c=-1.5$~a.u., $V_i=0.5$~a.u., and bath energies spanning
$\mu_{d,i}\in[-1.0,1.0]$~a.u. We obtained the exact spectral reference by
direct diagonalization and used the CCSD truncation for the real-time CC
calculations over $t\in[0,250]$~a.u. We converged the neutral ground-state
calculation to an energy change below $10^{-6}$~a.u. and an amplitude-norm
change below $10^{-7}$.

% ============================================================
\section{Supporting Theory for the \texorpdfstring{RT-$\Lambda$CC}{RT-LambdaCC} Formulation}
\label{sec:si_theory}
% ============================================================

This section provides additional details of the RT-$\Lambda$CC formulation
introduced in the main text. We focus on the correlated sudden-ionization
initial condition, the reduction of the $\Lambda$-weighted overlap, and the
relation of the resulting single-exponential propagation to
double-coupled-cluster (dCC) theory.

% ------------------------------------------------------------
\subsection{Correlated sudden-ionization state and shifted initialization}
\label{sec:si_shifted_initialization}
% ------------------------------------------------------------

The exact diagonal retarded single-particle Green's function for a core
spin-orbital \(c\) separates into electron-addition and electron-removal
components,
\begin{align*}
G_c^R(t)
&=
-i\Theta(t)
\langle\Psi_g^{(N)}|\{a_c(t),a_c^\dagger\}|\Psi_g^{(N)}\rangle,
\\
&=G_{c,+}^R(t)+G_{c,-}^R(t).
\end{align*}
The electron-removal component relevant to photoemission is exactly
\begin{equation}
G_{c,-}^R(t)
=
-i\Theta(t)e^{-iE_g^{(N)}t}
\langle\Psi_c^{(N-1)}|
e^{iHt}
|\Psi_c^{(N-1)}\rangle ,
\label{eq:si_gf_sudden}
\end{equation}
where the sudden core-hole state is
\begin{equation}
|\Psi_c^{(N-1)}\rangle
=
a_c|\Psi_g^{(N)}\rangle .
\label{eq:si_sudden_state}
\end{equation}
Because this study addresses the photoelectron spectrum, we retain
\(G_{c,-}^R(t)\) and denote it by \(G_c^R(t)\) below. Thus,
Eq.~\eqref{eq:si_gf_sudden} and Eq.~1 of the main article are the same exact
one-particle Green's-function formulation within the electron-removal sector.
The full retarded Green's function additionally contains the electron-addition
component. The subsequent coupled-cluster parametrization and CCSD truncation
introduce separate many-body approximations.
Within coupled-cluster theory,
\begin{align}
|\Psi_g^{(N)}\rangle
&=
e^{T^{(N)}}|\phi_0^{(N)}\rangle ,
\\
\langle\Psi_g^{(N)}|
&=
\langle\phi_0^{(N)}|
(1+\Lambda^{(N)})e^{-T^{(N)}} .
\end{align}
Because \(a_c\) commutes with the neutral excitation operator
\(T^{(N)}\),
\begin{equation}
|\Psi_c^{(N-1)}\rangle
=
e^{T^{(N)}}|\phi_c^{(N-1)}\rangle ,
\qquad
|\phi_c^{(N-1)}\rangle
=
a_c|\phi_0^{(N)}\rangle .
\label{eq:si_sudden_cc}
\end{equation}

We partition the neutral cluster operator as
\begin{equation}
T^{(N)}=T_r+T_c,
\label{eq:si_T_split}
\end{equation}
where \(T_r\) contains excitations that do not involve the core orbital
\(c\), while \(T_c\) contains excitations from \(c\). Since the core orbital
is unoccupied in \(|\phi_c^{(N-1)}\rangle\),
\begin{equation}
T_c|\phi_c^{(N-1)}\rangle=0,
\end{equation}
and the sudden state reduces to
\begin{equation}
|\Psi_c^{(N-1)}\rangle
=
e^{T_r}|\phi_c^{(N-1)}\rangle .
\label{eq:si_sudden_Tr}
\end{equation}
Thus, instantaneous core-electron removal preserves the non-core correlation
already present in the neutral ground state.

We parameterize the real-time equation-of-motion coupled-cluster
(RT-EOM-CC) propagation as
\begin{equation}
|\Psi_c^{(N-1)}(t)\rangle
=
N_c(t)e^{S_r(t)}|\phi_c^{(N-1)}\rangle ,
\label{eq:si_rt_ansatz}
\end{equation}
where \(S_r(t)\) contains excitation operators in the core-ionized
\((N-1)\)-electron space and therefore contains no operator that annihilates
the removed core orbital \(c\). Matching Eq.~\eqref{eq:si_rt_ansatz} to the
sudden state at \(t=0\) gives
\begin{equation}
S_r(0)=T_r.
\label{eq:si_shifted_ic}
\end{equation}
We therefore write
\begin{equation}
S_r(t)=T_r+\Delta S_r(t),
\qquad
\Delta S_r(0)=0.
\label{eq:si_delta_S}
\end{equation}
Shifting only the initial amplitudes from zero to \(T_r\) leaves the
single-exponential RT-EOM-CC equations of motion unchanged. The amplitudes
\(\Delta S_r(t)\) consequently describe changes
relative to the correlated sudden-ionization state.

% ------------------------------------------------------------
\subsection{\texorpdfstring{$\Lambda$-weighted}{Lambda-weighted} overlap}
\label{sec:si_lambda_overlap}
% ------------------------------------------------------------

We write the RT-$\Lambda$CC Green's function as
\begin{equation}
G_c^R(t)
\simeq
-i\Theta(t)e^{-iE_{\mathrm{CC}}^{(N)}t}
N_c(t)O_c^\Lambda(t).
\label{eq:si_G_lambda}
\end{equation}
For clarity, we denote the factor common to every overlap contribution by
\begin{equation}
P_c(t)
\equiv
-i\Theta(t)e^{-iE_{\mathrm{CC}}^{(N)}t}N_c(t),
\qquad
G_c^R(t)=P_c(t)O_c^\Lambda(t).
\label{eq:si_common_green_factor}
\end{equation}
Here, $P_c(t)$ carries the common phase and cumulant generated by the
core-hole propagation. We include it when transforming an overlap term to
the binding-energy domain because $O_2(t)$, $O_{3a}$, and $O_{3b}(t)$ are
components of the dimensionless overlap rather than separate Green's
functions. Multiplying each term by the same $P_c(t)$ places its Fourier
weight on the same binding-energy axis as the total spectrum without changing
the overlap-rank or fragment classification. The overlap factor itself is
\begin{equation}
O_c^\Lambda(t)
=
\langle\phi_0^{(N)}|
(1+\Lambda^{(N)})e^{-T^{(N)}}
a_c^\dagger e^{S_r(t)}
|\phi_c^{(N-1)}\rangle .
\label{eq:si_O_def}
\end{equation}
Here, \(\Lambda^{(N)}\) is the standard de-excitation operator defining the
left state in the biorthogonal coupled-cluster representation.

Using
\(|\phi_c^{(N-1)}\rangle=a_c|\phi_0^{(N)}\rangle\), together with
\begin{equation}
[a_c,T^{(N)}]=0,
\qquad
[a_c^\dagger,S_r(t)]=0,
\end{equation}
we express the overlap as
\begin{equation}
O_c^\Lambda(t)
=
\langle\phi_0^{(N)}|
(1+\Lambda^{(N)})e^{-T^{(N)}}
e^{S_r(t)}
a_c^\dagger a_c
|\phi_0^{(N)}\rangle .
\label{eq:si_O_projected}
\end{equation}
The number operator \(a_c^\dagger a_c\) projects onto configurations in
which \(c\) remains occupied. Consequently,
\begin{equation}
a_c^\dagger a_c
|\phi_0^{(N)}\rangle
=
|\phi_0^{(N)}\rangle .
\label{eq:si_projector_Tr}
\end{equation}

Using the shifted parametrization
\(S_r(t)=T_r+\Delta S_r(t)\), and noting that the relevant non-core
excitation operators commute, the static non-core contribution cancels
explicitly,
\begin{align}
e^{-T^{(N)}}e^{S_r(t)}
&=
e^{-(T_r+T_c)}
e^{T_r+\Delta S_r(t)}
\nonumber\\
&=
e^{-T_c}e^{-T_r}e^{T_r}e^{\Delta S_r(t)}
\nonumber\\
&=
e^{-T_c}e^{\Delta S_r(t)} .
\label{eq:si_Tr_cancellation}
\end{align}
The overlap therefore reduces to
\begin{equation}
O_c^\Lambda(t)
=
\langle\phi_0^{(N)}|
(1+\Lambda^{(N)})
e^{-T_c}
e^{\Delta S_r(t)}
|\phi_0^{(N)}\rangle .
\label{eq:si_O_working}
\end{equation}
Equation~\eqref{eq:si_O_working} separates the static neutral correlation
inherited by the sudden state from the additional response generated after
core ionization. The former determines the shifted initial condition,
whereas the latter enters through \(\Delta S_r(t)\) and the
\(\Lambda^{(N)}\)-weighted left-state overlap.

% ------------------------------------------------------------
\subsection{CCSD-level overlap contributions}
\label{sec:si_ccsd_overlap}
% ------------------------------------------------------------

At the CCSD level, we partition
\begin{equation}
\Lambda^{(N)}=\Lambda_r+\Lambda_c,
\end{equation}
where \(\Lambda_r\) contains no core index and \(\Lambda_c\) contains terms
involving \(c\). The overlap separates as
\begin{equation}
O_c^\Lambda(t)
=
1+O_2(t)+O_{3a}+O_{3b}(t).
\label{eq:si_O_partition}
\end{equation}
The non-core contribution is
\begin{align}
O_2(t)
&=
\sum_{ia}\lambda_a^i\Delta s_i^a(t)
+\frac{1}{4}\sum_{ijab}
\lambda_{ab}^{ij}\Delta s_{ij}^{ab}(t)
\nonumber\\
&\quad+
\frac{1}{4}\sum_{ijab}\lambda_{ab}^{ij}
\left[
\Delta s_i^a(t)\Delta s_j^b(t)
-
\Delta s_i^b(t)\Delta s_j^a(t)
\right],
\label{eq:si_O2}
\end{align}
where the occupied indices \(i,j\) exclude \(c\).

The static core-sector contribution is
\begin{equation}
O_{3a}
=
-\sum_a\lambda_a^c t_c^a
-\frac{1}{2}\sum_{iab}
\lambda_{ab}^{ic}t_{ic}^{ab},
\label{eq:si_O3a}
\end{equation}
and the time-dependent core-mediated contribution is
\begin{equation}
O_{3b}(t)
=
-\frac{1}{2}\sum_{iab}
\lambda_{ab}^{ic}
\left[
t_c^b\Delta s_i^a(t)
-
t_c^a\Delta s_i^b(t)
\right].
\label{eq:si_O3b}
\end{equation}
Thus, \(O_2(t)\) describes the non-core overlap response,
\(O_{3a}\) provides the static core-sector contribution, and
\(O_{3b}(t)\) couples the neutral core-sector amplitudes to the propagated
non-core response. We can evaluate these terms from the stored
RT-EOM-CC amplitudes without modifying the real-time equations of motion.

We checked the numerical hierarchy of these terms in both the time and
frequency domains. At the equilibrium water-dimer geometry, the
root-mean-square (RMS) magnitudes are
$\operatorname{RMS}|O_2|=6.41\times10^{-2}$,
$|O_{3a}|=2.53\times10^{-5}$, and
$\operatorname{RMS}|O_{3b}|=2.43\times10^{-7}$. Therefore, the time-dependent
core-mediated term has only $3.79\times10^{-6}$ of the RMS magnitude of
$O_2(t)$.  A small time-domain amplitude could in principle become relatively
visible near a zero of a larger Fourier component, so we also transformed
each contribution after multiplication by the common factor $P_c(t)$ in
Eq.~\eqref{eq:si_common_green_factor}. With the
same damping, finite record, and transform convention used for the spectrum,
the integrated absolute transform magnitudes over 550--575~eV, normalized by
that of the total response, are $2.00\times10^{-1}$ for
$P_c(t)O_2(t)$, $2.04\times10^{-5}$ for
$P_c(t)O_{3a}$, and $8.61\times10^{-8}$ for
$P_c(t)O_{3b}(t)$. Moreover, at frequencies where the total
satellite signal exceeds 1\% of its maximum, the combined $O_3$ transform is
never more than $2.75\times10^{-5}$ of the total.  The Fourier transform does
not reveal an appreciable hidden $O_3$ channel: $O_2(t)$ dominates the
nontrivial overlap correction. Plotting an independently normalized
$O_3$ transform would obscure this absolute hierarchy.

% ------------------------------------------------------------
\subsection{Relation to double-coupled-cluster theory}
\label{sec:si_dcc_relation}
% ------------------------------------------------------------

The correlated shifted-start construction is closely related to the double-exponential
structure of dCC~\cite{peng2024exploring,abraham2026elucidating}. To make the
comparison explicit, partition the neutral cluster operator as
$T^{(N)}=T_r+T_c$ and the additional ionized-sector operator as
$R(t)=R_r(t)+R_c(t)$. Here, $T_c$ contains neutral excitations from the core
orbital, while $R_c$ denotes ionized-sector terms whose contractions
explicitly involve the core-hole line (for example, refilling the core while
opening an additional valence hole). The remaining components, $T_r$ and
$R_r$, contain no core index. The schematic dCC ket is
\begin{equation}
e^{T^{(N)}}e^{R(t)}|\phi_c^{(N-1)}\rangle .
\label{eq:si_dcc_ket}
\end{equation}
Writing the product as a single exponential gives, through cubic order in the
cluster operators,
\begin{align}
e^{T^{(N)}}e^{R(t)}
&=e^{Z(t)},\nonumber\\
Z(t)
&=T^{(N)}+R(t)+\frac{1}{2}[T^{(N)},R(t)]\nonumber\\
&\quad+\frac{1}{12}[T^{(N)},[T^{(N)},R(t)]]\nonumber\\
&\quad+\frac{1}{12}[R(t),[R(t),T^{(N)}]]+\cdots .
\label{eq:si_bch}
\end{align}
Because the core-free excitation operators commute,
$[T_r,R_r(t)]=0$, their linear contribution combines directly as
$T_r+R_r(t)$. RT-$\Lambda$CC retains this part through the shifted
single-exponential amplitude
\begin{equation}
S_r(t)=T_r+\Delta S_r(t).
\label{eq:si_bch_retained}
\end{equation}

The first terms not reproduced by this shift are the mixed, second-order
Baker--Campbell--Hausdorff (BCH) commutators
\begin{equation}
\frac{1}{2}\left(
[T_c,R_r]+[T_r,R_c]+[T_c,R_c]
\right),
\label{eq:si_bch_second_order}
\end{equation}
where we suppress the time argument on $R$. At the next order, examples are
\begin{equation}
\begin{aligned}
&[T_r,[T_c,R_r]],\quad [T_c,[T_c,R_r]],\\
&[R_r,[R_r,T_c]],\quad [R_c,[T_r,R_c]],
\end{aligned}
\label{eq:si_bch_nested_examples}
\end{equation}
with the coefficients fixed by Eq.~\eqref{eq:si_bch}. Every omitted term
contains at least one core-sector operator, $T_c$ or $R_c$. Thus, the first
formal difference from dCC is already bilinear in a neutral core-sector and an
ionized-sector amplitude, whereas the nested commutators are cubic and higher
in the cluster amplitudes. With singles-and-doubles operators, their connected
contractions can generate effective higher-rank intermediates even though we
do not propagate an independent triple amplitude.

We can state the same distinction directly at the Hamiltonian level. The dCC
equations project the double-similarity-transformed Hamiltonian
\begin{align}
\overline{H}_T
&=e^{-T^{(N)}}He^{T^{(N)}},\nonumber\\
\overline{\overline{H}}_{\mathrm{dCC}}(t)
&=e^{-R(t)}\overline{H}_T e^{R(t)}\nonumber\\
&=\overline{H}_T+[\overline{H}_T,R]\nonumber\\
&\quad+\frac{1}{2}[[\overline{H}_T,R],R]+\cdots .
\label{eq:si_dcc_double_similarity}
\end{align}
After expanding $\overline{H}_T$ in $T_r+T_c$, this expression contains, for
example, $[[H,T_c],R_r]$,
$[[[H,T_r],T_c],R_r]/2$, and
$[[[H,T_c],R_r],R_r]/2$, together with counterparts involving $R_c$.
These connected terms feed neutral core-excitation information back into the
instantaneous ionized-sector residuals and therefore alter the propagated
amplitude equations. Physically, they describe hole-mediated pathways in
which core relaxation and valence screening occur in a coupled sequence,
rather than as a shifted initial condition followed by purely ionized-sector
relaxation.

RT-$\Lambda$CC does not evaluate these separate $T_c$- and $R_c$-dependent
Hamiltonian commutators. It imports the core-free neutral amplitudes through
Eq.~\eqref{eq:si_bch_retained} and includes core-sector information in the
$\Lambda^{(N)}$-weighted spectral overlap, but we evaluate that overlap after
propagation, so it does not feed back into the equations of motion. This is the
precise sense in which RT-$\Lambda$CC retains the leading dCC-like correlated
projection while omitting higher-order core-mediated dynamical couplings.

% ============================================================
\section{Comparison of Correlated Shifted-Start \texorpdfstring{RT-$\Lambda$CC}{RT-LambdaCC} with Double-Coupled-Cluster Propagation}
\label{sec:si_dcc_comparison}
% ============================================================

To assess how the shifted initialization and the $\Lambda$-weighted overlap
approximate the corresponding quantities in real-time dCC (RT-dCC) theory, we
compare the real-time quantities entering the
RT-$\Lambda$CC Green's function directly with RT-dCC.
Figure~2(b,c) of the main article compares the instantaneous core-hole energy
$E(t)$ and overlap function $O(t)$ from conventional real-time
coupled-cluster (RT-CC), correlated shifted-start RT-$\Lambda$CC, and RT-dCC.
Because the main article already presents the figure, we provide below the
additional theoretical context needed to interpret that comparison.

The correlated shifted-start RT-$\Lambda$CC propagation closely follows RT-dCC in
both the real and imaginary parts of $E(t)$, whereas conventional RT-CC
shows appreciable deviations. Because the RT-$\Lambda$CC equations of
motion are otherwise unchanged from RT-CC, this improvement arises from
initializing the non-core amplitudes with the correlation inherited from
the neutral sudden-ionization state. The comparison therefore shows that
the shifted initial condition reproduces much of the effect of the
neutral-sector correlation on the core-hole propagation for this
benchmark.

The overlap provides a complementary test. Conventional RT-CC uses a
trivial overlap, $O(t)=1$, and therefore does not retain the
time-dependent projection onto the correlated neutral CC left state.
The $\Lambda$-weighted overlap in RT-$\Lambda$CC instead closely follows
the RT-dCC overlap in both its real and imaginary components. Thus, the
two ingredients of RT-$\Lambda$CC address distinct aspects of the
double-CC description: the shifted initialization improves the
core-hole propagation, while the $\Lambda$-weighted overlap restores
the nontrivial left-state projection entering the Green's function.

For the present benchmark, RT-$\Lambda$CC therefore reproduces the
principal real-time behavior of RT-dCC without introducing the
double-exponential propagation or the associated core-mediated BCH
terms into the equations of motion. The remaining differences reflect
dynamical couplings retained in RT-dCC but omitted from the
single-exponential RT-$\Lambda$CC propagation.

% ============================================================
\section{Fragment-Resolved Orbital Analysis}
\label{sec:si_fragment_basis}
% ============================================================

We propagated the RT-$\Lambda$CC dynamics in the canonical restricted
Hartree--Fock (RHF) molecular-orbital basis. To interpret the overlap
components in terms of local and intermolecular excitation channels, we
constructed a fragment-canonical orbital basis as a post-processing
transformation. Previous real-time time-dependent density functional theory
(TDDFT) studies of intermolecular Coulombic decay have used related
fragment-resolved orbital analyses~\cite{wang2024nonhermitian_icd}.

For the water dimer, we kept the two O~$1s$ core orbitals separate and
localized the occupied valence and virtual spaces independently using
Pipek--Mezey (PM) rotations,
\begin{align}
C_{\mathrm{val}}^{\mathrm{loc}}
&=
C_{\mathrm{val}}U_{\mathrm{val}}^{\mathrm{PM}},
\\
C_{\mathrm{vir}}^{\mathrm{loc}}
&=
C_{\mathrm{vir}}U_{\mathrm{vir}}^{\mathrm{PM}}.
\end{align}
We assigned each localized orbital $\phi_p$ to fragment
$X\in\{A,B\}$ according to its fragment population,
\begin{equation}
P_X(\phi_p)
=
\sum_{\mu\in X}
c_{\mu p}(S c_p)_\mu ,
\label{eq:si_fragment_population}
\end{equation}
where $S$ is the atomic-orbital (AO) overlap matrix and $c_p$ is the AO coefficient
vector of orbital $p$.

\begin{figure}
\centering
\includegraphics[width=0.8\linewidth]{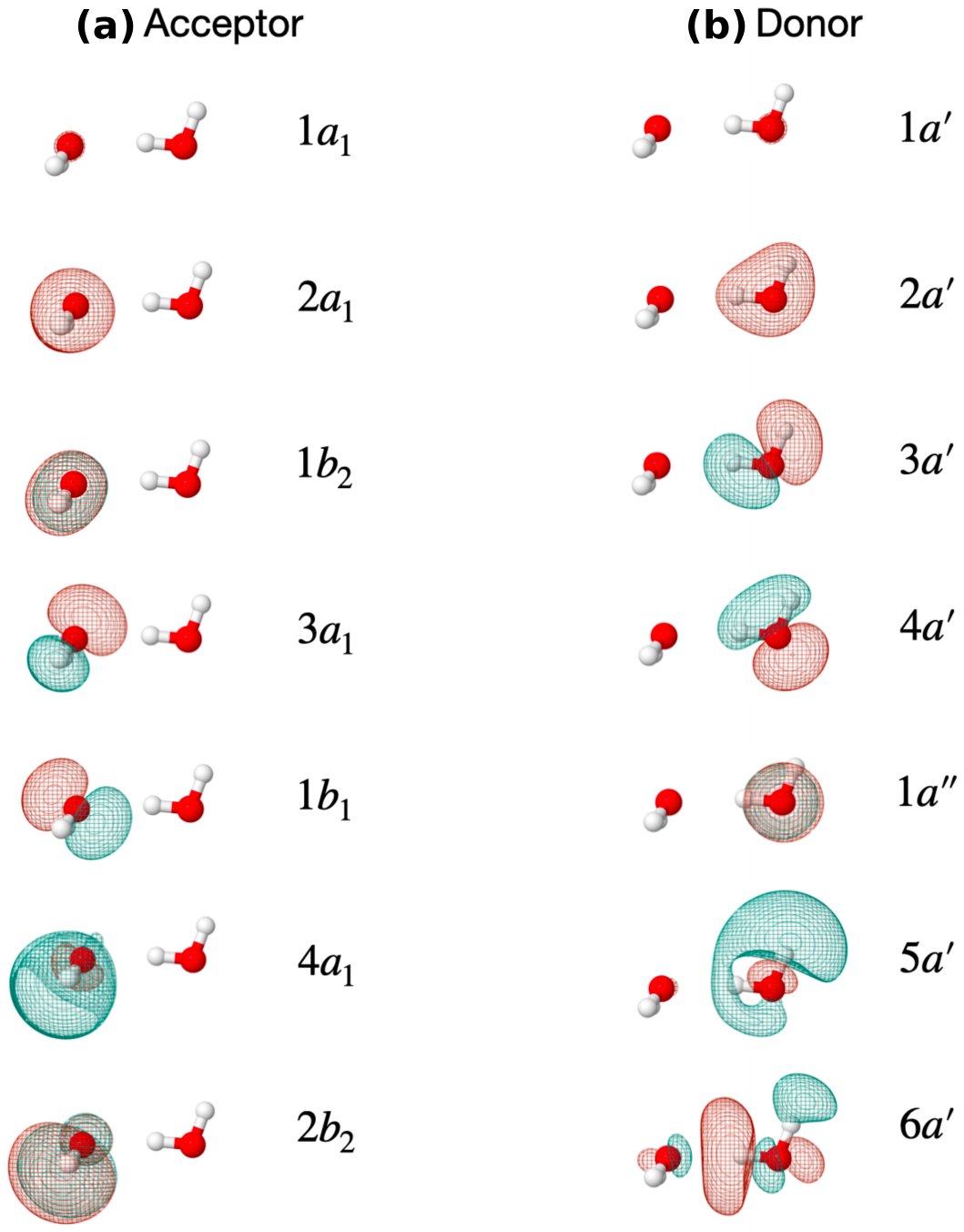}
\caption{Fragment-canonical orbitals of the water dimer used for the
orbital- and fragment-resolved component analysis. We keep the two O~$1s$
core orbitals separate and assign the valence occupied and virtual orbitals
to the acceptor ($A$) and donor ($B$) fragments.}
\label{fig:water_dimer_mo}
\end{figure}

To obtain orbitals with well-defined fragment character and an energy
ordering within each fragment, we subsequently canonicalized the localized
orbitals within each fragment block. We diagonalized the projected Fock matrix
\begin{equation}
F_X
=
(C_X^{\mathrm{loc}})^\dagger
F^{\mathrm{AO}}
C_X^{\mathrm{loc}}
\end{equation}
as
\begin{equation}
F_XV_X=V_X\varepsilon_X,
\end{equation}
yielding the fragment-canonical orbitals
\begin{equation}
C_X^{\mathrm{fc}}
=
C_X^{\mathrm{loc}}V_X.
\end{equation}
We applied this procedure separately to the occupied valence and virtual
subspaces of fragments $A$ and $B$. Figure~\ref{fig:water_dimer_mo} shows the
corresponding orbitals and retains the donor $a'/a''$ labels. In the same
order, $1a'$, $2a'$, $3a'$, $4a'$, $1a''$, $5a'$, and $6a'$ map to the
B:$1a_1$, B:$2a_1$, B:$1b_2$, B:$3a_1$, B:$1b_1$, B:$4a_1$, and B:$2b_2$
labels used throughout the analysis.

We did not repeat the real-time propagation in the fragment-canonical
basis. Instead, we transformed the stored canonical-basis overlap components
into the fragment-canonical representation using the
corresponding unitary occupied- and virtual-space rotations. For example,
the singles amplitudes transform as
\begin{equation}
\widetilde{s}_i^{\,a}(t)
=
\sum_{pq}
(U_{\mathrm{occ}}^\dagger)_{ip}
s_p^{\,q}(t)
(U_{\mathrm{vir}})_{qa},
\label{eq:si_singles_rotation}
\end{equation}
and applied analogous transformations to the doubles and left-state tensors.

Because these transformations are unitary within the occupied and virtual
subspaces, they leave the total overlap invariant,
\begin{equation}
\widetilde{O}_c^\Lambda(t)
=
O_c^\Lambda(t),
\label{eq:si_overlap_invariance}
\end{equation}
while redistributing its individual tensor components among orbitals with
well-defined fragment character. We can therefore classify the transformed
components as $A$-local, $B$-local, or charge-transfer-like
according to the fragment labels of the occupied and virtual orbitals.
This transformation changes only the representation used for component
analysis and does not alter the propagated RT-$\Lambda$CC dynamics or the
total spectrum.

% ============================================================
\section{Orbital-Resolved CC Excitation Weight and Time--Frequency Analysis}
\label{sec:si_orbital_weight}
% ============================================================

To identify which occupied-orbital-indexed amplitude groups change following
core ionization, we analyze the propagated RT-$\Lambda$CC amplitudes relative
to the correlated sudden-ionization state,
\begin{equation}
\Delta S_r(t)=S_r(t)-S_r(0)=S_r(t)-T_r .
\label{eq:si_deltaS_weight}
\end{equation}
For an occupied fragment-canonical orbital \(p\), we define the
orbital-resolved CC excitation weight in the spin-resolved CCSD
representation as
\begin{align}
H_p(t)
&=
\sum_a\left|\Delta s_{p_\alpha}^{a_\alpha}(t)\right|^2
+\sum_a\left|\Delta s_{p_\beta}^{a_\beta}(t)\right|^2
\nonumber\\
&\quad+
\sum_{jab}
\left|\Delta s_{p_\alpha j_\beta}^{a_\alpha b_\beta}(t)\right|^2
+\sum_{iab}
\left|\Delta s_{i_\alpha p_\beta}^{a_\alpha b_\beta}(t)\right|^2.
\label{eq:si_orbital_weight}
\end{align}
The indices \(a,b\) are virtual orbitals, and \(i,j\) are the remaining
occupied valence orbitals. Equation~\eqref{eq:si_orbital_weight} expands the
compact main-text expression by counting each listed occurrence of \(p\) once,
which defines the multiplicity factor \(w_{\mu p}\).

The CC amplitudes are dynamical coordinates of the propagated coupled-cluster
state. An amplitude containing occupied index \(p\) therefore contributes to
the group labeled by orbital \(p\). For a double excitation carrying occupied
indices \(p\) and \(q\), the same amplitude contributes to both \(H_p\) and
\(H_q\). The orbital-indexed groups are consequently overlapping projections,
not mutually exclusive channels that can be added to form a population.
Similar bands in acceptor and donor panels therefore do not pair those orbitals.
Using \(\Delta S_r(t)\), rather than \(S_r(t)\), removes the neutral-state
correlation inherited through \(S_r(0)=T_r\), so \(H_p(0)=0\) and the
diagnostic isolates amplitude changes generated after core-hole creation.

The quantity \(H_p(t)\) is an amplitude-based diagnostic rather than an
observable. It is not an orbital occupation, density, charge population, or
expectation value, and its magnitude does not measure electrons removed from
or transferred between orbitals or fragments. Because the decomposition also
depends on the orbital representation, we report it only in the
fragment-canonical representation defined in
Sec.~\ref{sec:si_fragment_basis}.

To resolve when this excitation weight develops and its characteristic time
scales, we apply the STFT,
\begin{equation}
\widetilde H_p(\omega,\tau)
=
\int dt\,W(t-\tau)H_p(t)e^{i\omega t},
\label{eq:si_stft}
\end{equation}
where \(W(t-\tau)\) is a finite-time window centered at \(\tau\).
For independent regeneration, we fit and subtract a quartic polynomial from
each archived $H_p(t)$ trace. On the 0.1~a.u. output grid, we then apply a
symmetric 827-sample Hann window, corresponding to 1.998~fs, and advance its
center by 103 samples, corresponding to 0.249~fs. Fourfold zero padding gives
a 3308-point transform and a displayed Fourier grid spacing of 0.517~eV. The
approximately 2~fs window gives the nominal Fourier energy scale
$h/T_{\mathrm{win}}\approx2.1$~eV. The smaller center increment samples the
time dependence more finely but does not improve this intrinsic energy
resolution. At every geometry, the finite B:$3a_1$ trace extends to 9.499~fs
and the 31 STFT window centers span 0.999--8.473~fs. Figure~5(a) of the main
article uses these independently regenerated arrays on a common time axis
without extrapolation or time-axis stretching.
The plotted STFT magnitude is
\begin{equation}
A_p(\omega,\tau)=\left|\widetilde H_p(\omega,\tau)\right|,
\label{eq:si_stft_magnitude}
\end{equation}
which we use as a relative time--frequency weight. Its
square, \(F_p(\omega,\tau)=A_p(\omega,\tau)^2\), is the corresponding
spectrogram power. Without another nonlinear transformation, we normalize each
composite by one common reference $A_{\mathrm{ref}}$, the 99.5th percentile
over its panels, and cap the scale at 1. We do not normalize panels
individually. With the same window and figure-wide normalization, cross-panel and within-map
time--energy comparisons are valid only as relative modulation comparisons.
The STFT is a local frequency decomposition and does not require
\(\Delta s_\mu(t)\) to be a single exponentially decaying function. Complex
exponentials serve only as a Fourier basis. A non-exponential build-up or slow
envelope contributes predominantly at low frequency, whereas rapid curvature,
an abrupt onset, and finite-window leakage can distribute weight over a broader
frequency range.

\begin{figure*}[!t]
  \centering
  \includegraphics[width=0.95\textwidth]{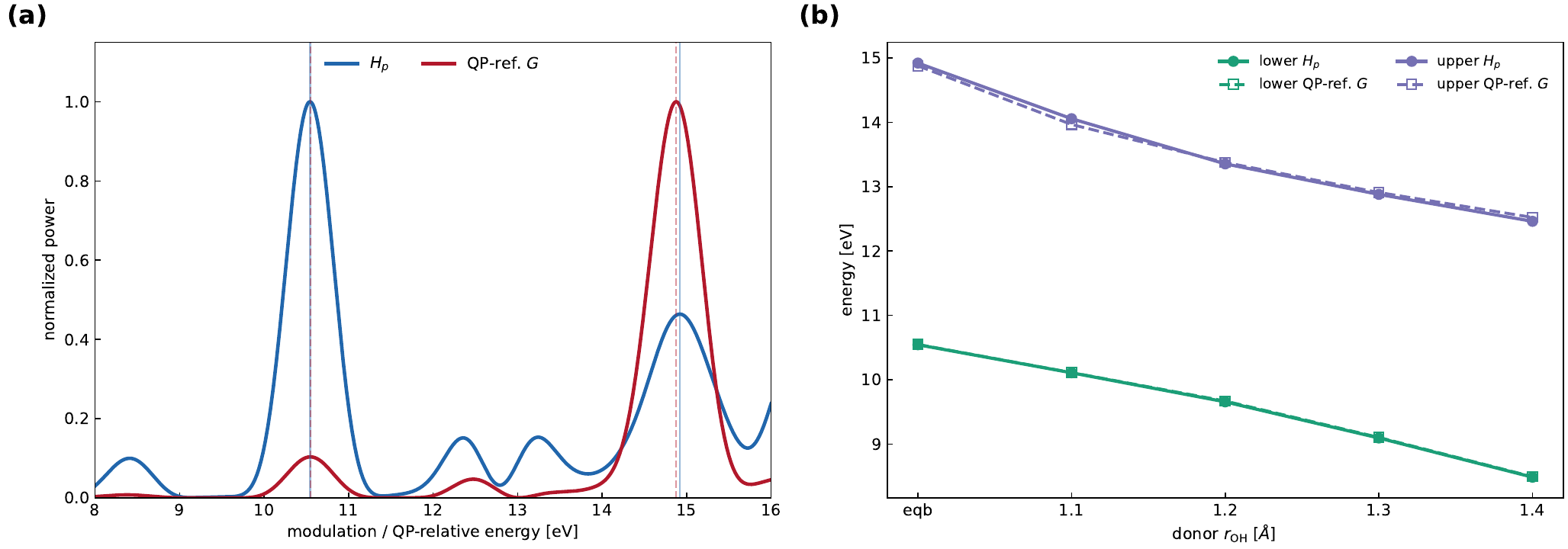}
  \caption{QP-referenced interpretation of the donor B:$3a_1$
  excitation-weight modulation. (a) Equilibrium power spectra of the
  quartic-detrended $H_{B:3a_1}(t)$ trace and the QP-referenced residual
  $g_{\mathrm{rel}}(t)$. We normalize each spectrum independently over the
  displayed 8--16~eV interval. Solid blue and dashed red vertical lines mark
  the paired local maxima of the $H_p$ and QP-referenced signals,
  respectively. (b) Evolution of two tracked local-maximum trajectories along
  the donor $r_{\mathrm{OH}}$ scan. ``Lower'' and ``upper'' designate the two
  $H_p$ maxima ordered by energy and followed continuously between neighboring
  geometries. They do not denote electronic-state eigenvalue branches. Filled
  circles denote $H_p$ modulation maxima and transparent squares denote the
  corresponding QP-relative features of $G(t)$. Their covariation supports an
  important QP--satellite beating contribution to $H_p(t)$ but does not provide
  a unique state-by-state assignment.
  Table~\ref{tab:si_qp_coherence} reports the quantitative coherences.}
  \label{fig:si_h_qp_alignment}
\end{figure*}

Figure~\ref{fig:si_h_qp_alignment} summarizes the empirical frequency
alignment and local-maximum tracking relative to the quasiparticle (QP),
thereby motivating the interpretation and coherence test below.

To distinguish the STFT and X-ray photoelectron spectroscopy (XPS) energy
axes, we first consider the exact
stationary-state expansion of the sudden core-ionized wave packet,
\begin{equation}
a_c|\Psi_0^N\rangle
=\sum_K d_K|\Psi_K^{N-1}\rangle,
\label{eq:si_ionized_expansion}
\end{equation}
where \(E_K^{N-1}\) is the energy of ionized state \(K\). The removal part of
the exact core-hole Green's function has the Lehmann form
\begin{equation}
G_c(t)=-i\Theta(t)\sum_K Z_K
e^{-iI_Kt/\hbar},
\qquad
I_K=E_K^{N-1}-E_0^N,
\label{eq:si_xps_lehmann}
\end{equation}
where \(Z_K=|d_K|^2\) in the exact Hermitian theory. Fourier transforming
Eq.~\eqref{eq:si_xps_lehmann} therefore produces XPS poles directly at the
binding energies \(I_K\).

We apply the STFT to a different, quadratic quantity. If, within a finite
analysis window, we can represent an RT-CC amplitude approximately by modal
components associated with the ionized-state energies,
\begin{equation}
\Delta s_\mu(t)\simeq
\sum_K b_{\mu K}e^{-i\omega_Kt},
\qquad
\hbar(\omega_K-\omega_L)=I_K-I_L,
\label{eq:si_amplitude_modal}
\end{equation}
where a common phase offset in all \(\omega_K\) is immaterial,
then its squared magnitude contains bilinear terms
\begin{equation}
|\Delta s_\mu(t)|^2\simeq
\sum_{KL}b_{\mu K}b_{\mu L}^{*}
e^{-i(\omega_K-\omega_L)t}.
\label{eq:si_amplitude_bilinear}
\end{equation}
The corresponding modulation energies are therefore
\begin{equation}
E_{\mathrm{mod}}^{KL}=\hbar|\Omega_{KL}|=|I_K-I_L|.
\label{eq:si_modulation_difference}
\end{equation}
In the special case that \(L\) is the quasiparticle state,
\begin{equation}
E_{\mathrm{mod}}^{K,\mathrm{QP}}
=I_K-I_{\mathrm{QP}}
\equiv\Delta E_K,
\label{eq:si_qp_satellite_gap}
\end{equation}
so a modulation band can coincide with a satellite separation from the main
line. If both \(K\) and \(L\) are satellites, however, the same construction
produces a satellite--satellite energy difference rather than the position of
either satellite. Multiple state pairs can also share the same difference
energy.

Equations~\eqref{eq:si_amplitude_modal}--\eqref{eq:si_qp_satellite_gap} are a
conditional interpretation rather than an exact identification of the RT-CC
amplitudes with stationary-state coefficients. The amplitudes obey coupled
nonlinear equations and are representation-dependent coordinates of the CC
state. A rigorous stationary-state analysis requires explicit biorthogonal
EOM-CC or response-theory projectors~\cite{pedersen2021interpretation}.
Nonlinear combination frequencies and transient/windowing contributions may
therefore accompany the idealized difference frequencies in
Eq.~\eqref{eq:si_modulation_difference}.

The archived numerical traces permit a more direct QP-referenced coherence
test. The historical archive key `activity' contains the raw $H_p(t)$
excitation-weight traces, while `stft\_values' contains the traces used for the
frequency analysis.
A fourth-order polynomial reproduces their difference with a maximum error
below $2\times10^{-15}$. The valid B:$3a_1$ traces extend to 9.499~fs at
all five geometries and have a 0.1~a.u. output interval. Over 5--32~eV, the
cosine similarity between the raw and quartic-detrended Hann-windowed power
spectra is greater than 0.99999 at every geometry, showing that the bands used
below are insensitive to removal of the slow background.

For the coherence test, we located the QP energy $I_{\mathrm{QP}}$ from the
damped transform of $G(t)$ using $\eta=0.01$~a.u. We then removed a
least-squares QP exponential and referenced the residual to the QP phase,
schematically,
\begin{equation}
g_{\mathrm{rel}}(t)=
e^{iI_{\mathrm{QP}}t/\hbar}
\left[G(t)-a_{\mathrm{QP}}e^{-iI_{\mathrm{QP}}t/\hbar}\right].
\label{eq:si_qp_referenced_signal}
\end{equation}
The signs in the numerical implementation follow the stored Green's-function
convention. Equation~\eqref{eq:si_qp_referenced_signal} states the physical
frequency shift.  We evaluated the magnitude-squared coherence
\begin{equation}
C_p(E)=
\frac{\left|\left\langle
\widetilde h_p(E,\tau)\widetilde g_{\mathrm{rel}}^{*}(E,\tau)
\right\rangle_{\tau}\right|^2}
{\left\langle|\widetilde h_p(E,\tau)|^2\right\rangle_{\tau}
 \left\langle|\widetilde g_{\mathrm{rel}}(E,\tau)|^2\right\rangle_{\tau}},
\label{eq:si_qp_coherence}
\end{equation}
using 2~fs Hann segments advanced by 0.25~fs. Because zero padding only
interpolates the displayed frequency grid, it does not alter the nominal
$\sim2.1$~eV resolution.

Table~\ref{tab:si_qp_coherence} quantifies the paired modulation energies and
coherences used in this comparison.

\begin{table}[t]
  \centering
  \caption{QP-referenced coherence test for the two strongest low-energy
  B:$3a_1$ tracked modulation maxima. Each lower- or upper-energy entry is
  $E[H_{B:3a_1}]/E[g_{\mathrm{rel}}]$ in eV, and $C$ is the corresponding
  magnitude-squared coherence.}
  \label{tab:si_qp_coherence}
  \scriptsize
  \setlength{\tabcolsep}{2pt}
  \begin{tabular*}{0.98\columnwidth}{@{\extracolsep{\fill}}lcccc@{}}
    \toprule
    $r_{\mathrm{OH}}$ & \shortstack{Lower\\energy} & $C_{\mathrm{lower}}$ &
    \shortstack{Upper\\energy} & $C_{\mathrm{upper}}$ \\
    \midrule
    eqb (0.964~\AA) & 10.543/10.548 & 0.945 & 14.920/14.877 & 0.347 \\
    1.1~\AA          & 10.108/10.111 & 0.959 & 14.056/13.966 & 0.525 \\
    1.2~\AA          &  9.656/ 9.669 & 0.950 & 13.354/13.375 & 0.871 \\
    1.3~\AA          &  9.094/ 9.104 & 0.896 & 12.879/12.910 & 0.953 \\
    1.4~\AA          &  8.486/ 8.493 & 0.928 & 12.461/12.522 & 0.968 \\
    \bottomrule
  \end{tabular*}
\end{table}

The maximum mismatch between paired numerical maxima is 0.090~eV. Because
this value lies well below the intrinsic STFT resolution, it does not
constitute an independent sub-eV energy determination. The significant observation
is instead the covariation of both tracked maxima with geometry and their
coherence: the lower maximum remains strongly coherent throughout the scan,
while the upper maximum evolves from weak coherence at equilibrium to strong coherence
upon O--H elongation. Although this covariation supports a substantial
QP--satellite bilinear contribution to $H_{B:3a_1}(t)$, satellite--satellite,
nonlinear, and transient terms account for additional structure.

The broad equilibrium B:\(3a_1\) band also contains the independently tabulated
QP-relative satellite separations 10.683, 12.820, 14.957, and 15.811~eV whose
ranked motifs involve B:\(3a_1\). The A:\(1b_1\) and A:\(1b_2\) bands
similarly coincide, within the nominal Fourier resolution, with the
21.366/22.221 and 25.212/25.640~eV satellite separations carrying the same
occupied-orbital labels.  Accordingly, the STFT and coherence analyses provide
a time--frequency consistency check on the correlated amplitude response, not
a separate photoelectron spectrum or a measure of time-dependent charge
transfer.

Figure~4 of the main article shows the equilibrium multi-orbital STFT maps, so
we do not repeat them here. The definitions above specify the plotted quantity
and delimit its interpretation.

\section{Intermolecular-Separation Control}
\label{sec:si_oo_separation}
% ============================================================

To test whether the nonlocal satellite character arises from
intermolecular coupling, we varied the O$\cdots$O separation while keeping the
intramolecular geometries and relative orientation of the two water
molecules fixed. We considered the equilibrium separation
\(R_{\mathrm{OO}}=2.91~\text{\AA}\), a shortened separation of
\(2.70~\text{\AA}\), and a well-separated geometry at
\(5.00~\text{\AA}\).

The low-energy satellite structure remains similar at
\(R_{\mathrm{OO}}=2.70\) and \(2.91~\text{\AA}\), whereas separating the
monomers to \(5.00~\text{\AA}\) strongly suppresses the \(B\)-local features
(Figure~3(a) of the main article). We do not repeat the spectrum here. The
fragment-resolved component areas show the same behavior
(Table~\ref{tab:oo-separation-fragment-character}).

\begin{table}[t]
\centering
\footnotesize
\caption{Fragment character of the satellite-region component area as a
function of intermolecular O$\cdots$O separation. We obtain the fractions by
integrating the absolute component area over the satellite window and
normalizing by the corresponding total satellite component area, excluding
the O~1s quasiparticle contribution.}
\label{tab:oo-separation-fragment-character}
\begin{tabular*}{0.98\columnwidth}{@{\extracolsep{\fill}}lccc@{}}
\toprule
\(R_{\mathrm{OO}}\) & \(A\)-local & \(B\)-local & CT \\
\midrule
\(2.70~\text{\AA}\) & 0.735 & 0.033 & 0.232 \\
\(2.91~\text{\AA}\) & 0.775 & 0.027 & 0.198 \\
\(5.00~\text{\AA}\) & 0.962 & 0.006 & 0.031 \\
\bottomrule
\end{tabular*}
\end{table}

At \(5.00~\text{\AA}\), the response is almost entirely \(A\)-local
(0.962), with both the \(B\)-local (0.006) and CT (0.031) contributions
strongly reduced. At the hydrogen-bonded separations, the nonlocal
contributions are substantially larger, while changing
\(R_{\mathrm{OO}}\) from 2.91 to \(2.70~\text{\AA}\) produces only a
modest additional redistribution. Because we leave the intramolecular
geometries unchanged in this comparison, the disappearance of the low-energy
\(B\)-local features at large separation supports their assignment to
intermolecular coupling rather than an isolated-monomer response.

% ============================================================
\section{Basis-Set Robustness}
\label{sec:si_basis}
% ============================================================

To assess the sensitivity of the RT-$\Lambda$CCSD results to the
one-electron basis, we compare calculations using the 6-31++G** and
aug-cc-pVDZ basis sets at both the equilibrium and stretched water-dimer
geometries. We examine both the total O~$1s$ spectrum and the
fragment-resolved composition of the satellite region.

\begin{figure*}[t]
    \centering
    \includegraphics[width=0.72\textwidth]{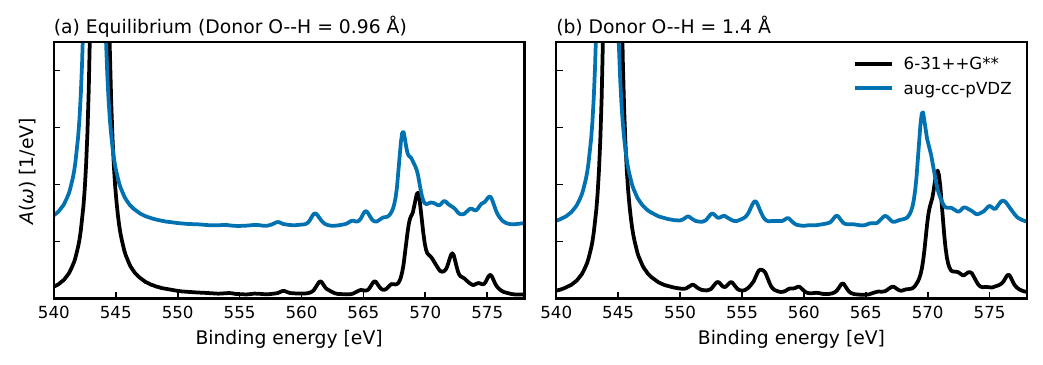}
    \caption{Comparison of the RT-$\Lambda$CCSD O~$1s$ spectra computed
    with the 6-31++G** and aug-cc-pVDZ basis sets at
    (a) the equilibrium geometry, with donor O--H
    $r_{\mathrm{OH}}=0.964~\text{\AA}$, and
    (b) the stretched geometry, with
    $r_{\mathrm{OH}}=1.4~\text{\AA}$. Within each panel, the two curves use a
    common vertical scale and the aug-cc-pVDZ curve is shifted upward for
    clarity. We generate the spectra from the archived time responses with
    $\eta=0.01$~a.u., a raised-cosine taper over the final 25\% of each time
    record, fourfold Fourier zero padding, and 0.1~eV Gaussian smoothing.}
    \label{fig:basis-comparison-eqb-r14}
\end{figure*}

Figure~\ref{fig:basis-comparison-eqb-r14} shows that the two basis sets
produce closely similar spectral profiles at both geometries. In
particular, aug-cc-pVDZ reproduces the principal satellite features at
equilibrium and their pronounced reorganization upon O--H elongation. The
remaining differences are primarily quantitative changes
in individual satellite positions and intensities.

To test whether the fragment character underlying this spectral
reorganization is similarly robust, Table~\ref{tab:fragment-character-basis}
compares the integrated $A$-local, $B$-local, and charge-transfer-like
contributions. We used the same component construction and satellite
integration procedure for both basis sets.

\begin{table}[t]
\centering
\scriptsize
\caption{Basis robustness of the fragment character of the satellite-region component area for
the equilibrium and stretched water-dimer geometries. We obtain the fractions
by integrating the absolute component area over the satellite window and
normalizing by the corresponding total satellite component
area, excluding the O~1s quasiparticle contribution.}
\label{tab:fragment-character-basis}
\setlength{\tabcolsep}{2.8pt}
\begin{tabular*}{0.98\columnwidth}{@{\extracolsep{\fill}}llccc@{}}
\toprule
Geometry & Basis & $A$-local & $B$-local & CT \\
\midrule
Equilibrium & 6-31++G**    & 0.775 & 0.027 & 0.198 \\
Equilibrium & aug-cc-pVDZ  & 0.711 & 0.043 & 0.246 \\
$r_{\mathrm{OH}}=1.4~\text{\AA}$ & 6-31++G**   & 0.536 & 0.148 & 0.315 \\
$r_{\mathrm{OH}}=1.4~\text{\AA}$ & aug-cc-pVDZ & 0.451 & 0.197 & 0.352 \\
\bottomrule
\end{tabular*}
\end{table}

Although the absolute component fractions depend somewhat on the basis,
both calculations give the same qualitative redistribution. At
equilibrium, the classified satellite component area is predominantly
$A$-local, with
smaller CT and $B$-local contributions. Upon stretching to
$r_{\mathrm{OH}}=1.4~\text{\AA}$, the $A$-local fraction decreases
substantially, while both the $B$-local and CT fractions increase. For
6-31++G**, the $B$-local contribution increases from 0.027 to 0.148 and
the CT contribution from 0.198 to 0.315. The aug-cc-pVDZ calculation gives the
same trend,
with increases from 0.043 to 0.197 and from 0.246 to 0.352, respectively.

Thus, both the spectral reorganization and the accompanying increase in
nonlocal fragment character upon donor O--H elongation remain qualitatively
robust when we change from 6-31++G** to aug-cc-pVDZ. This comparison tests the
robustness of the physical trend rather than complete basis-set convergence.

%============================================================
\section{Peak-Level Satellite Assignments and Comparison with Monomer Symmetry-Adapted-Cluster Configuration Interaction (SAC-CI)}
\label{sec:si_sacci}
% ============================================================

Table~2 of the main article reports the full set of principal ranked motifs
used to support the three broad spectral regimes. We do not duplicate that
table here. The motifs identify the dominant ranked components and do not
constitute a complete state-by-state decomposition.

We compare the higher-energy $A$-local satellite features with the O~$1s$
shake-up states that Sankari \emph{et al.} previously calculated for an
isolated water molecule using SAC-CI~\cite{Sankari2006SACCI}. Because the
SAC-CI states contain mixtures of several configurations, we use the
comparison to establish correspondence in energy scale and dominant orbital
character rather than a one-to-one assignment of individual states.
Table~\ref{tab:h2odimer_sacci_comparison} summarizes the comparison.

\begin{table}[!htbp]
\centering
\scriptsize
\caption{Comparison of selected $A$-local satellite features of the
equilibrium water dimer with water-monomer SAC-CI results from
Ref.~\cite{Sankari2006SACCI}. We report energies in eV relative to the
corresponding O~$1s$ main line.}
\label{tab:h2odimer_sacci_comparison}
\setlength{\tabcolsep}{2.4pt}
\begin{tabular*}{0.98\columnwidth}{@{\extracolsep{\fill}}cccc@{}}
\toprule
\shortstack{Present\\$\Delta E$} & \shortstack{Present\\character} &
\shortstack{SAC-CI\\$\Delta E$} & \shortstack{SAC-CI\\character} \\
\midrule
17.95 & $3a_1$-type & 18.19 & $3a_1$-type \\
21.37 & $1b_1$-type & 20.66 & $1b_1$-type \\
22.22 & $1b_1$-type & 23.37 & $1b_1$-type \\
\bottomrule
\end{tabular*}
\end{table}

The correspondence in relative energy and dominant orbital character
supports the assignment of the higher-energy $A$-local dimer satellites
as predominantly monomer-like shake-up excitations. The comparison is
qualitative because hydrogen bonding modifies the dimer states and the
two calculations employ different correlated electronic-structure
treatments.

\FloatBarrier

\clearpage
\bibliographystyle{apsrev4-2}
\bibliography{library}

\end{document}